\documentclass[5p, preprint]{elsarticle}
\makeatletter 
\def\ps@pprintTitle{%
 \let\@oddhead\@empty
 \let\@evenhead\@empty
 \let\@evenfoot\@oddfoot}
\makeatother
\usepackage[utf8]{inputenc}
\usepackage[T1]{fontenc}
\usepackage[english]{babel}
\usepackage[babel=true]{csquotes}
\usepackage[fleqn]{amsmath} 
\usepackage{amsthm} 
\usepackage{booktabs}
\usepackage{multirow} 
\usepackage{amssymb} 
\usepackage{float}
\usepackage{hyperref} 
\usepackage[english]{cleveref} 
\usepackage{subcaption}
\usepackage{caption}
\usepackage{natbib}
\usepackage{soul}
\usepackage{color}
\usepackage[dvipsnames,table]{xcolor}
\usepackage{pifont}
\usepackage{graphicx}
\usepackage{svg}
\usepackage[inline]{enumitem}
\usepackage{array}
\usepackage{longtable}
\usepackage{tabularx}
\usepackage{graphicx}
\usepackage{makecell}
\usepackage{lscape}
\usepackage{threeparttable}
\usepackage{xurl}
\usepackage{wasysym}
\usepackage[shortcuts]{extdash}

\definecolor{ao}{rgb}{0.0, 0.5, 0.0}
\newcommand{\yestick}{{\color{ao}\ding{51}}}
\newcommand{\notick}{{\color{red}\ding{55}}}

\newtoggle{finalPaper}
\toggletrue{finalPaper} 

\iftoggle{finalPaper} {

    \newcommand{\rmvtxt}[1]{}
} {

    \newcommand{\rmvtxt}[1]{\st{#1}}
}

\definecolor{gray45}{gray}{.45}
\definecolor{gray75}{gray}{.75}
\definecolor{orange-fig}{HTML}{C55A11}

\journal{Information Fusion}

\biboptions{numbers}

\begin{document}

\begin{frontmatter}

\title{Data Fusion in Neuromarketing: Multimodal Analysis of Biosignals, Lifecycle Stages, Current Advances, Datasets, Trends, and Challenges}

\author[1]{Mario Quiles P\'erez\corref{cor1}}\ead{mqp@um.es}
\author[1]{Enrique Tom\'as Mart\'inez Beltr\'an}\ead{enriquetomas@um.es}
\author[1]{Sergio L\'opez Bernal}\ead{slopez@um.es}
\author[2]{Eduardo Horna Prat}\ead{eduardohorna@bitbrain.com}
\author[2]{Luis Montesano Del Campo}\ead{luis.montesano@bitbrain.es}
\author[3]{Lorenzo Fern\'andez Maim\'o}\ead{lfmaimo@um.es}
\author[4]{Alberto Huertas Celdr\'an}\ead{huertas@ifi.uzh.ch}

\cortext[cor1]{Corresponding author.}

\address[1]{Department of Information and Communications Engineering, University of Murcia, Murcia, 30100 Spain}
\address[2]{Bit\&Brain Technologies S.L, Zaragoza, 50006 Spain}
\address[3]{Department of Computer Architecture and Technology, University of Murcia, Murcia, 30100 Spain}
\address[4]{Communication Systems Group CSG, Department of Informatics IfI, University of Zurich UZH, CH---8050 Zürich, Switzerland}

\begin{abstract}

The primary goal of any company is to increase its profits by improving both the quality of its products and how they are advertised. In this context, neuromarketing seeks to enhance the promotion of products and generate a greater acceptance on potential buyers. Traditionally, neuromarketing studies have relied on a single biosignal to obtain feedback from presented stimuli. However, thanks to new devices and technological advances studying this area of knowledge, recent trends indicate a shift towards the fusion of diverse biosignals. An example is the usage of electroencephalography for understanding the impact of an advertisement at the neural level and visual tracking to identify the stimuli that induce such impacts. This emerging pattern determines which biosignals to employ for achieving specific neuromarketing objectives.
Furthermore, the fusion of data from multiple sources demands advanced processing methodologies. Despite these complexities, there is a lack of literature that adequately collates and organizes the various data sources and the applied processing techniques for the research objectives pursued. To address these challenges, the current paper conducts a comprehensive analysis of the objectives, biosignals, and data processing techniques employed in neuromarketing research. This study provides both the technical definition and a graphical distribution of the elements under revision.
Additionally, it presents a categorization based on research objectives and provides an overview of the combinatory methodologies employed. After this, the paper examines primary public datasets designed for neuromarketing research together with others whose main purpose is not neuromarketing, but can be used for this matter. Ultimately, this work provides a historical perspective on the evolution of techniques across various phases over recent years and enumerates key lessons learned.
\end{abstract}


\end{frontmatter}

\section{Introduction}
Neuromarketing is the discipline that applies neuropsychology to marketing research with the aim of studying and predicting the human behaviors generated by marketing practices. Historically, this area has been of interest to companies because it improves advertising campaigns and increases profits. In the last years, the techniques employed for obtaining user feedback have evolved towards the acquisition of biosignals \cite{kaheh2021study} and the automatic detection of mental states using Artificial Intelligence (AI) \cite{mouammine2019using}. In this context, more affordable and portable devices have been developed to acquire diverse biosignals. Brain-Computer Interfaces (BCIs) \cite{van2009brain}, Electrodermal Activity (EDA) sensors \cite{boucsein2012electrodermal}, or Eye-Tracking (ET) glasses \cite{tanenhaus1996eye} are well-known and representative examples of new devices used in neuromarketing to capture biological signals. These devices ease to conduct experiments based on the application of stimuli such as images or videos to a group of subjects and monitor their biosignals without needing professional equipment. Moreover, they provide multimodal data that are combined to yield more comprehensive insights \cite{goncalves2022neuromarketing}. Once the data are obtained from the subjects participating in the experimentation, the latest works propose the use of solutions based on AI to identify specific biosignal patterns that can be related to a mental state or reaction \cite{kumar2019fusion}. Traditionally, experts have performed this identification by gathering statistics from the signals and visually interpreting them \cite{cosic2016neuromarketing}. However, this has disadvantages, such as the subjectivity introduced when evaluating the results.

\begin{figure*}[!ht]
  \centering
  \includegraphics[width=\textwidth]{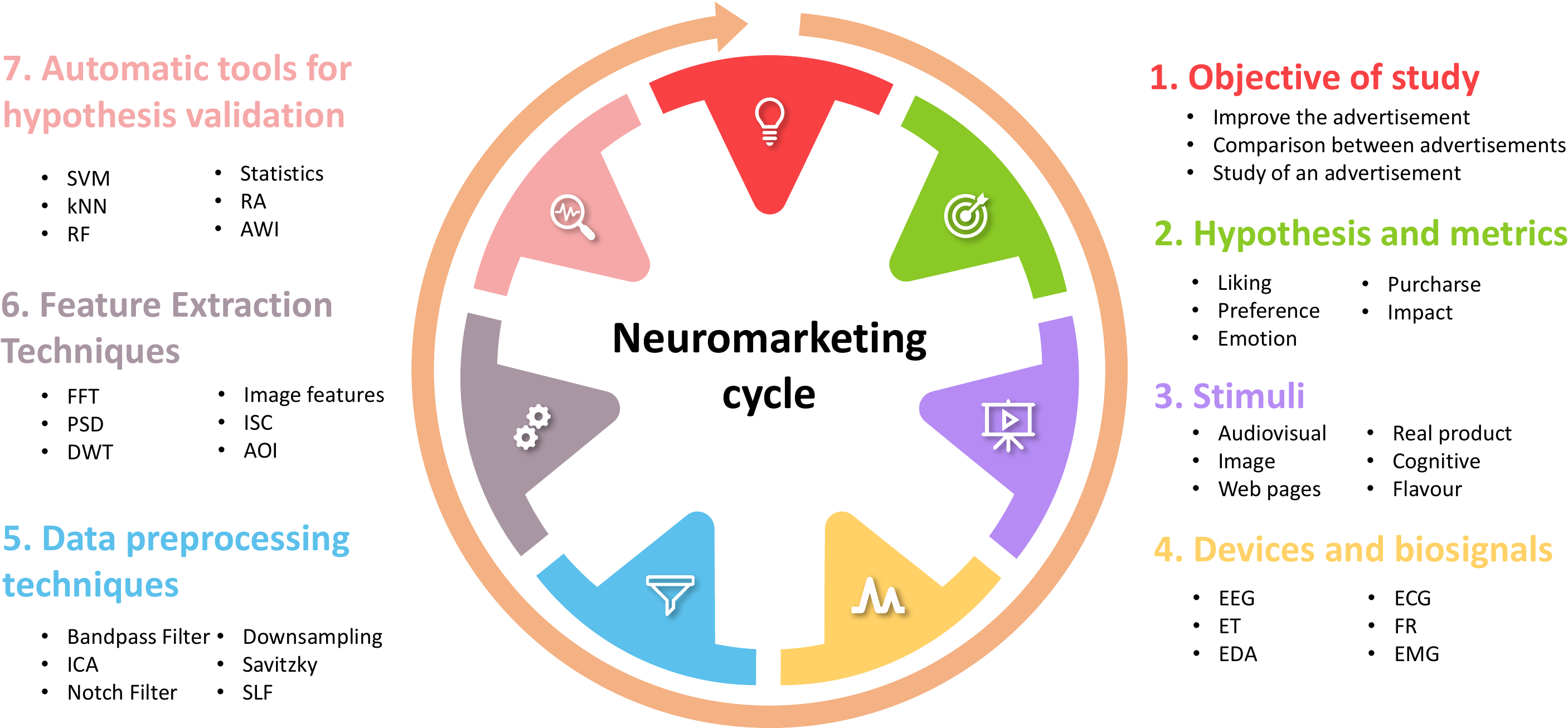}
  \caption{Neuromarketing cycle.}
  \label{fig:general}
\end{figure*}

Neuromarketing studies follow a life cycle composed of several stages to detect mental states from diverse biosignals. This cycle begins with the  definition of the study objective and ends with the validation of this objective (see \figurename~\ref{fig:general}) \cite{bazzani2020eeg}. In the first stage of the cycle, the following three possible objectives have been identified in the literature: improving the impact of an advertisement, comparing similar advertisements, and studying the impact of an advertisement
\cite{kumar2015neuromarketing, dragolea2011neuromarketing}. Once the objective is selected, the second stage defines the hypothesis together with the metrics used to measure the impact of the stimulus, namely, the emotional state evoked or product preference. This impact will be measured by the data obtained from the biosignals, so the relationship with phase four is very close. The third phase of the cycle decides the stimuli used to perform the study \cite{hakim2019gateway}, such as audiovisual (e.g., a video), or cognitive inputs, like browsing a web page. Next, during the fourth stage, the process determine which set of devices is the most appropriate to capture the biosignals of interest, for example a BCI, or an ET system. This device will provide a data flow of measurements. The fifth and sixth phases preprocess this data flow to clean the signal and extract relevant information. The algorithms applied in these two phases are key to a correct subsequent data fusion. The concluding phase involves data classification, which traditionally has been performed manually to yield statistics. However, the recent introduction of Machine Learning (ML) and Deep Learning (DL) algorithms into this field enables pattern recognition and reduces the subjectivity introduced by human judgment, further highlighting the tools used for hypothesis validation in the context of multimodal data fusion.

Each of the phases of neuromarketing studies hosts a wide range of alternatives. Therefore, the potential combinations increase exponentially, and finding the optimum combination to obtain the best performance becomes challenging and time-consuming. Moreover, there is no work in the literature that takes into account all phases of the neuromarketing cycle and explores the options at each stage. To tackle this challenge, the present work compiles the most representative efforts to raise similarities and differences in this context. Specifically, Bercea \cite{bercea2012anatomy} conducted the first literature review to highlight the advantages and disadvantages of the different biosignals used for neuromarketing objectives. Three years later, Dos et al., \cite{dos2015eye} studied the feasibility of using ET biosignals in real scenarios such as product selection or product pricing. After that, Mileti et al., \cite{mileti2016nanomarketing} presented the concept of nanomarketing from a theoretical point of view, analyzing how this concept could help current neuromarketing. Subsequently, due to the emergence of BCIs and EEG, and their application for mental state detection, the number of studies using this biosignal increased and, therefore, the number of surveys collecting information about them. Furthermore, Hakim et al., \cite{hakim2019gateway} and Bazzani et al., \cite{bazzani2020eeg} performed a compilation of works on the EEG signal applied to neuromarketing scenarios. During the year 2021, two more papers were published. The first one, by Kalaganis et al., \cite{kalaganis2021unlocking} presented, from a theoretical point of view, the different biosensors used for neuromarketing and in which scenarios they are used. In contrast, Khurana et al., \cite{khurana2021survey} studied those technical aspects necessary for the experiments, such as preprocessing, feature extraction, and data classification.

\begin{table*}[!ht]
  \caption{Surveys on neuromarketing and biosignals.}
  \label{tab:suveys}
  \resizebox{\textwidth}{!}{
  \begin{tabular}{cccccccccccc}
    \toprule
    \textbf{Ref} & \textbf{Year} & \textbf{Cites} & \textbf{Biosignal} & \textbf{Objectives} & \textbf{Stimuli} & \textbf{Preprocessing} & \begin{tabular}[c]{@{}c@{}}\textbf{Feature} \\ \textbf{Extraction}\end{tabular} & \textbf{Tools} & \textbf{Results} & \textbf{Datasets} & \begin{tabular}[c]{@{}c@{}}\textbf{Challenges} \\ \textbf{and Trends}\end{tabular} \\
    \midrule
    \midrule
    \cite{bercea2012anatomy} & 2012 & 130 & \begin{tabular}[c]{@{}c@{}}EEG, ECG, ETC, \\ FR, EDA, EMG\end{tabular}    & \notick & \notick & \notick & \notick & \notick & \notick & \notick & \notick \\ \midrule
    \cite{dos2015eye}& 2015& 103 & ET & \yestick & \yestick & \notick & \notick & \notick & \notick & \notick &  \notick \\ \midrule
    \cite{mileti2016nanomarketing} & 2016 & 15 & None & \notick & \notick & \notick & \notick & \notick & \notick &\notick & \yestick\\\midrule
    \cite{hakim2019gateway}& 2018 & 42 & EEG & \yestick & \yestick & \yestick & \yestick & \yestick & \notick & \notick & \yestick \\\midrule
    \cite{bazzani2020eeg}& 2020 & 16 & EEG & \yestick & \notick & \yestick & \notick & \notick & \notick & \notick & \yestick \\\midrule
    \cite{kalaganis2021unlocking}& 2021 & 1 & \begin{tabular}[c]{@{}c@{}}EEG, ECG, ETC, \\ FR, EDA, EMG\end{tabular}  & \yestick & \yestick & \notick & \notick & \notick & \notick & \notick & \notick \\\midrule
    \cite{khurana2021survey}& 2021 & 5 & EEG & \yestick & \yestick & \yestick & \yestick & \yestick & \notick & \yestick & \yestick  \\\midrule
    This work& 2023 & 0 & \begin{tabular}[c]{@{}c@{}}EEG, ECG, ETC, \\ FR, EDA, EMG\end{tabular}  & \yestick & \yestick & \yestick & \yestick & \yestick & \yestick & \yestick & \yestick  \\
    \bottomrule
    \bottomrule
  \end{tabular}}
\end{table*}

\tablename~\ref{tab:suveys} summarizes the previous works, showing the number of biosignals studied, as well as each of the phases of the neuromarketing cycle, and whether they are detailed or not. As a conclusion of the works described above, some challenges are still open. First, papers do not address the whole neuromarketing cycle, usually not including the technical part, such as data preprocessing, feature extraction, or data classification. Because of this, the reproducibility of the experiments becomes more complex, as there are many unknowns regarding the procedures to be performed to obtain the best results. Oppositely, those papers that detail the technical part of the work only consider a biosignal, usually the EEG signal. Another point not addressed in the literature are the available datasets that exist for this scenario. The lack of datasets makes it difficult to compare different methodologies proposed due to the dependence on the data captured. Thus, by addressing this aspect, a direct comparison could be made. Finally, the literature does not show the trends that each work has followed in recent years or the challenges that neuromarketing research will face in the coming years. Because of these shortcomings, this paper highlights the following research questions:

\begin{itemize}
\item {\verb|Q1|}: \textit{What are the most applied objectives, biosensors, and processing techniques in neuromarketing?} Specific biosensors and particular techniques are used depending on the objective of the campaign. There is a lack of works in the literature covering all these considerations.

\item {\verb|Q2|}: \textit{What combinations of data from biosignals, stimuli, and data processing and classification techniques are suitable for each objective?} Once all the techniques that have been applied for these investigations are known, the combinations between them, depending on the objective pursued, should be studied. Since no studies have been found considering all the campaign phases and the biosensors available, these associations are not reflected in the present works.

\item {\verb|Q3|}: \textit{What datasets are publicly available in neuromarketing, and what type of data include?} This study aims to identify existing public datasets that facilitate data fusion and are applicable to neuromarketing objectives. In this way, the problem of the dependence on the data obtained from the biosignals can be solved when comparing the methodologies proposed by the different works.

\item {\verb|Q4|}: \textit{How has the role of data fusion in neuromarketing evolved over time, and what are the open challenges?} Current surveys do not show the evolution of the work in terms of the objectives in neuromarketing, as well as the data fusion used over the years. Likewise, they do not show the evolution of the datasets that have appeared in recent years and for which objectives they have been proposed. Finally, an overview of the published works, limitations, and opportunities for further work needs to be identified.

\end{itemize}

\begin{figure*}[!ht]
  \centering
  \includegraphics[width=0.8\textwidth]{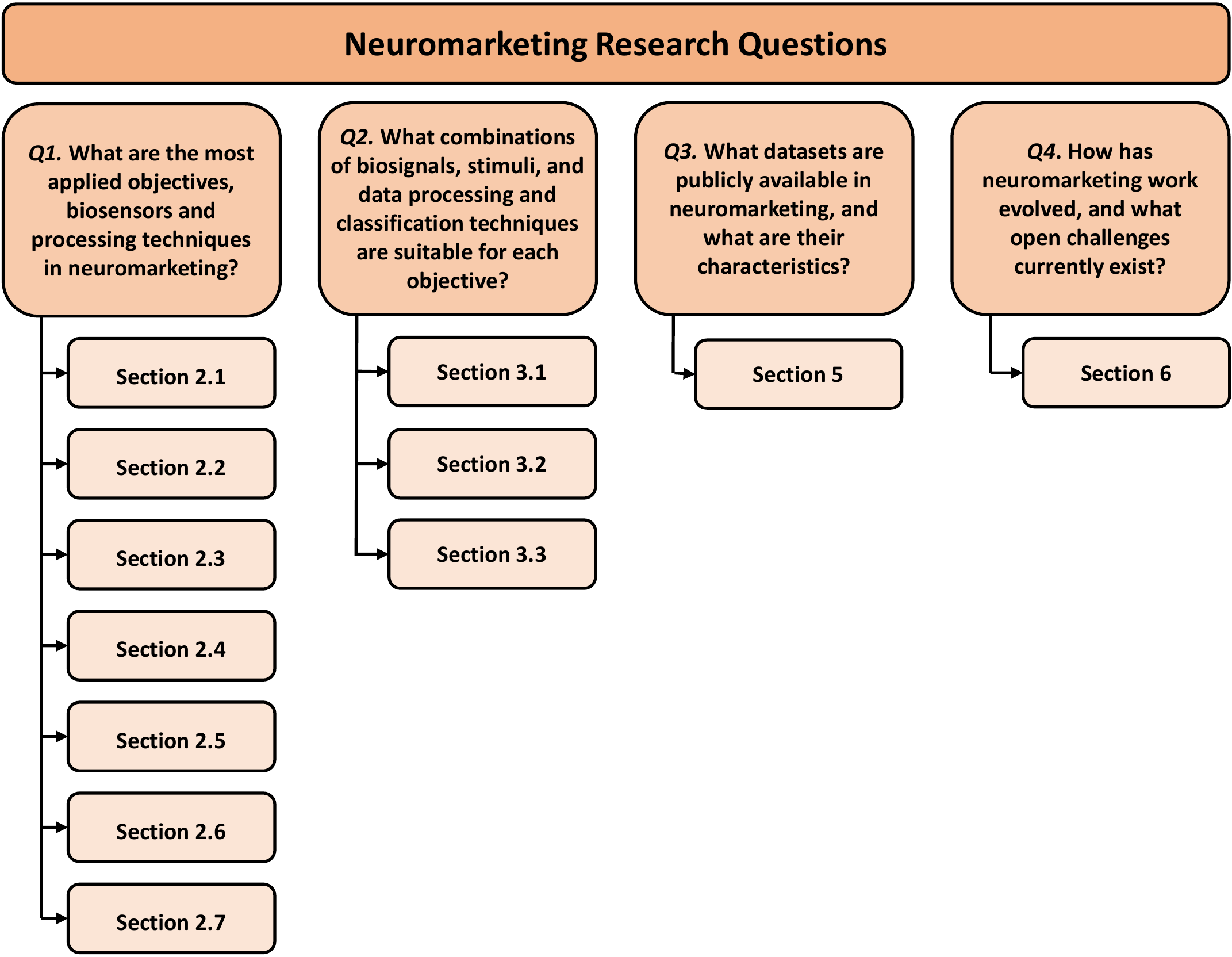}
  \caption{Discussed questions per article section.}
  \label{fig:questions}
\end{figure*}

To answer the above research questions, the main contributions of this paper are the following:

\begin{itemize}
    \item A complete analysis of the objectives, biosensors, and processing data techniques employed in neuromarketing. This study presents the technical definition and a graphical distribution of the use of the elements in question. This contribution responds to question \textit{Q1} in Section \ref{lifecycle}.
    
    \item A literature analysis based on neuromarketing, obtaining the main characteristics of each work applied in the neuromarketing cycle and grouped by objectives. In this sense, this paper presents in Section \ref{comparison} the combinations of techniques used in each of the works, as well as the results obtained, answering \textit{Q2}.
    
    \item An analysis of the primary public datasets explicitly designed for research work related to neuromarketing. This review answers question \textit{Q3}, detailing in Section \ref{datasets} those datasets focused on neuromarketing, together with others whose main purpose is not neuromarketing, but can be used for this purpose.
    
    \item An overview of how the techniques of the different phases have evolved over the last few years and a list of lessons learned, aiming to respond to question \textit{Q4} in Section \ref{ctl}.
    
\end{itemize}

The previous questions are interrelated and provide a comprehensive overview of the current state of the data and techniques used in neuromarketing. \textit{Q1} focus on stu\-/dying the objectives, biosensors, and techniques currently used in this type of research. In contrast, \textit{Q2} and \textit{Q3} show the combinations of data used in current literature and the existing datasets for prospecting work. These aspects represent the most critical sections of this paper. Finally, \textit{Q4} highlights the trends followed in recent years and the consequences of these trends. As a summary, \figurename~\ref{fig:questions} shows which sections of this paper answer the research questions presented.

The remainder of this article is organized as follows. Section \ref{lifecycle} analyzes biosensor types, objectives, and techniques used in neuromarketing. After that, Section \ref{comparison} describes and compares the leading solutions found in the state of the art. Moreover, Section \ref{datasets} examines the primary public datasets containing biosignals. Then, Section \ref{ctl} draws a set of lessons learned, current trends, and future challenges in the research area. Finally, Section \ref{conclusions} provides an insight into the conclusions extracted from the present work.

\section{What are the most applied objectives, biosensors and processing techniques in neuromarketing?}
\label{lifecycle}
This section answers the first research question by following a search methodology composed of the next steps.

\begin{itemize}
\item Definition of the subsequent search terms: Neuromarketing, Electroencephalography (EEG), EDA, Galvanic Skin Response (GSR), Electromyography (EMG), Electrocardiography (ECG), Eye-Tracker (ET), Face Recognition (FR), Biosignal, Survey, Review. These terms have been used in combinations of three or more words.
\item Selection of the scientific databases where the search for articles is performed: Google, Google Scholar, and Scopus.
\item Read the abstract and introduction of the resulting publications to determine if they fit the neuromarketing topic studied in this work.
\end{itemize}

\begin{figure*}[!ht]
  \centering
  \includegraphics[width=0.9\textwidth]{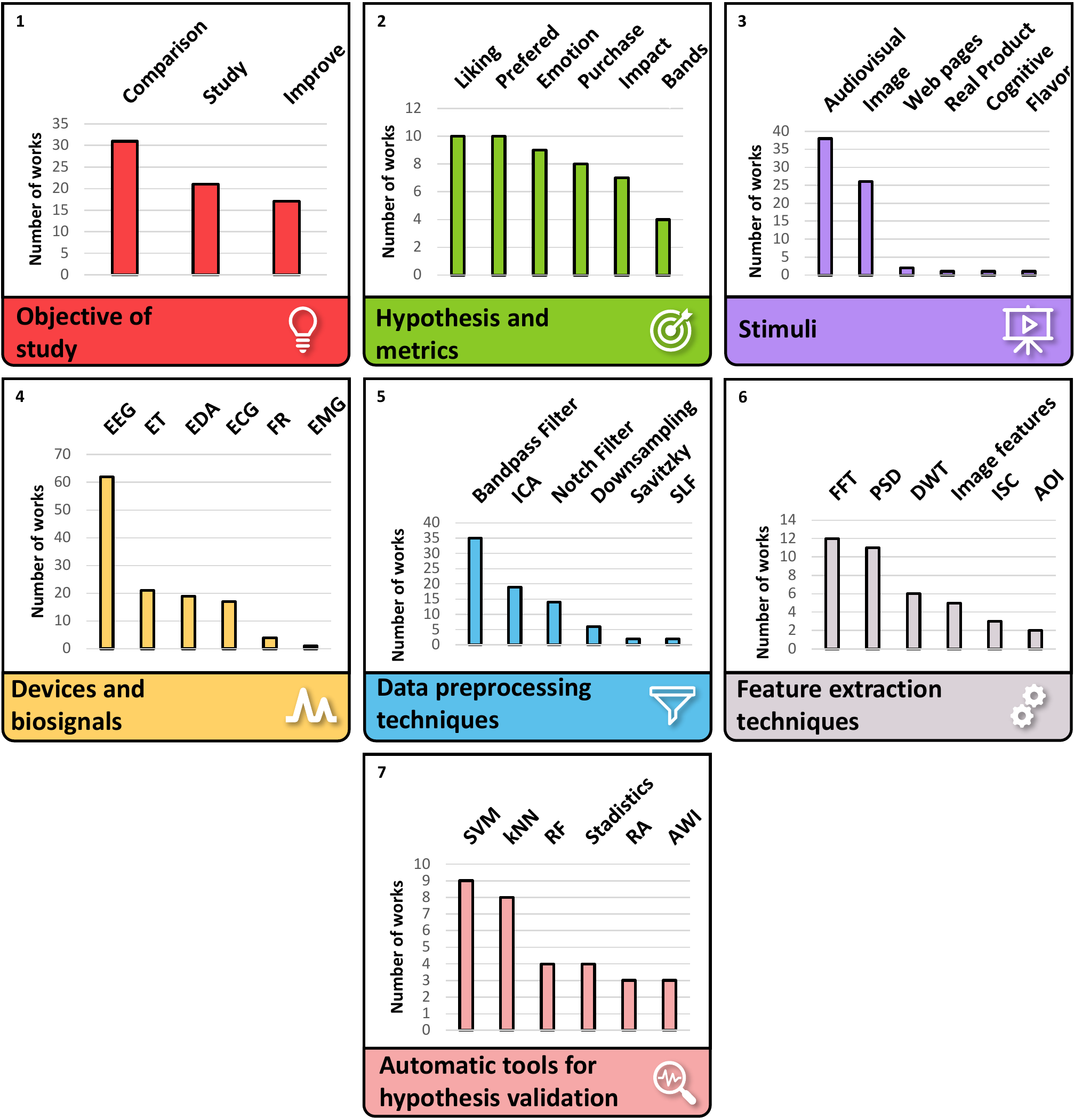}
  \caption{Number of works using the main techniques in each phase of the neuromarketing cycle.}
  \label{fig:lifecycle}
\end{figure*}

Through the use of this methodology, this work collects the most common neuromarketing phases among the studied papers \cite{shahriari2020meta}, being as follows (see \figurename~\ref{fig:lifecycle}):

\begin{enumerate}
    \item \textbf{Objective of study}: The first phase defines the objective of the study designed. Three objectives have been identified, (i) study the impact of a stimulus, (ii) modify the stimulus to improve its impact, and (iii) compare a stimulus with other similar ones.

    \item \textbf{Hypothesis and metrics}: The second phase focuses on the hypothesis posed to solve the objective and the metrics used to obtain a conclusion.
        
    \item \textbf{Stimuli}: The third phase focuses on the stimuli applied to subjects during the experimental period.
    
    \item \textbf{Devices and biosignals}: The fourth phase involves selecting both the biosignals used in the study and the biosensors needed to capture them.
    
    \item \textbf{Data processing techniques}: In order to eliminate possible noises introduced during acquisition, the analyzed works apply a data preprocessing phase.

    \item \textbf{Feature extraction techniques}: To extract the relevant information from the data, a feature extraction phase is applied.

    \item \textbf{Automatic tools for hypothesis validation}: Once the necessary information is available, the last phase is the classification stage. This phase studies the information coming from previous steps, and decides to which label it corresponds.

\end{enumerate}

The details of each phase are shown in Sections \ref{objectives}, \ref{hypothesis}, \ref{stimulus}, \ref{devices}, \ref{preprocessing}, \ref{features}, and  \ref{classification}. Moreover, \figurename~\ref{fig:lifecycle} shows the distribution charts between for the different stages. The graphs show the number of publications using each of the defined techniques and approaches.

\begin{figure*}[!ht]
    \begin{subfigure}{0.49\textwidth}
     \centering
     \includegraphics[width=\columnwidth]{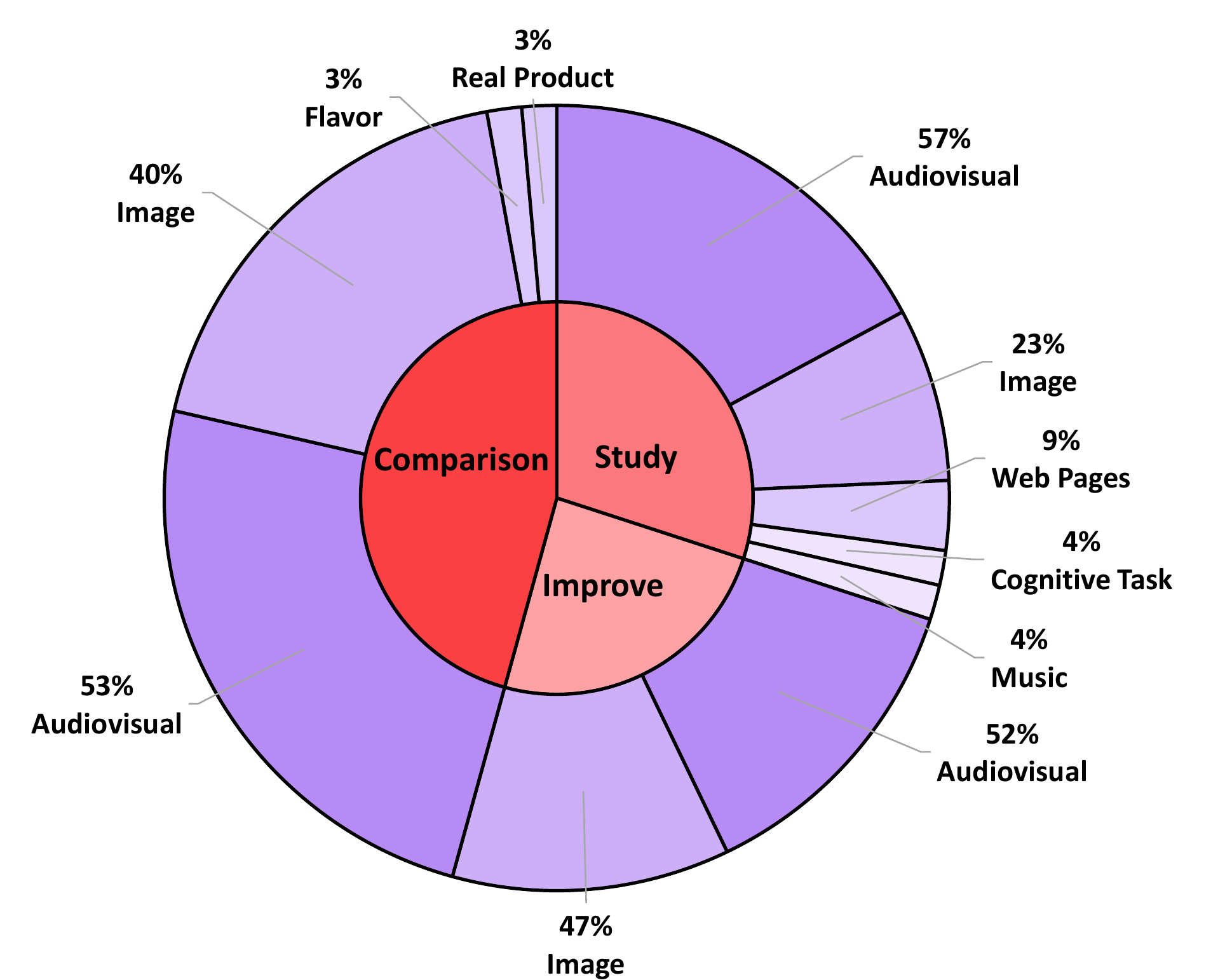}
     \caption{Stimulus.}
     \label{fig:disstimuli}
    \end{subfigure}
    \begin{subfigure}{0.49\textwidth}
     \centering
     \includegraphics[width=\columnwidth]{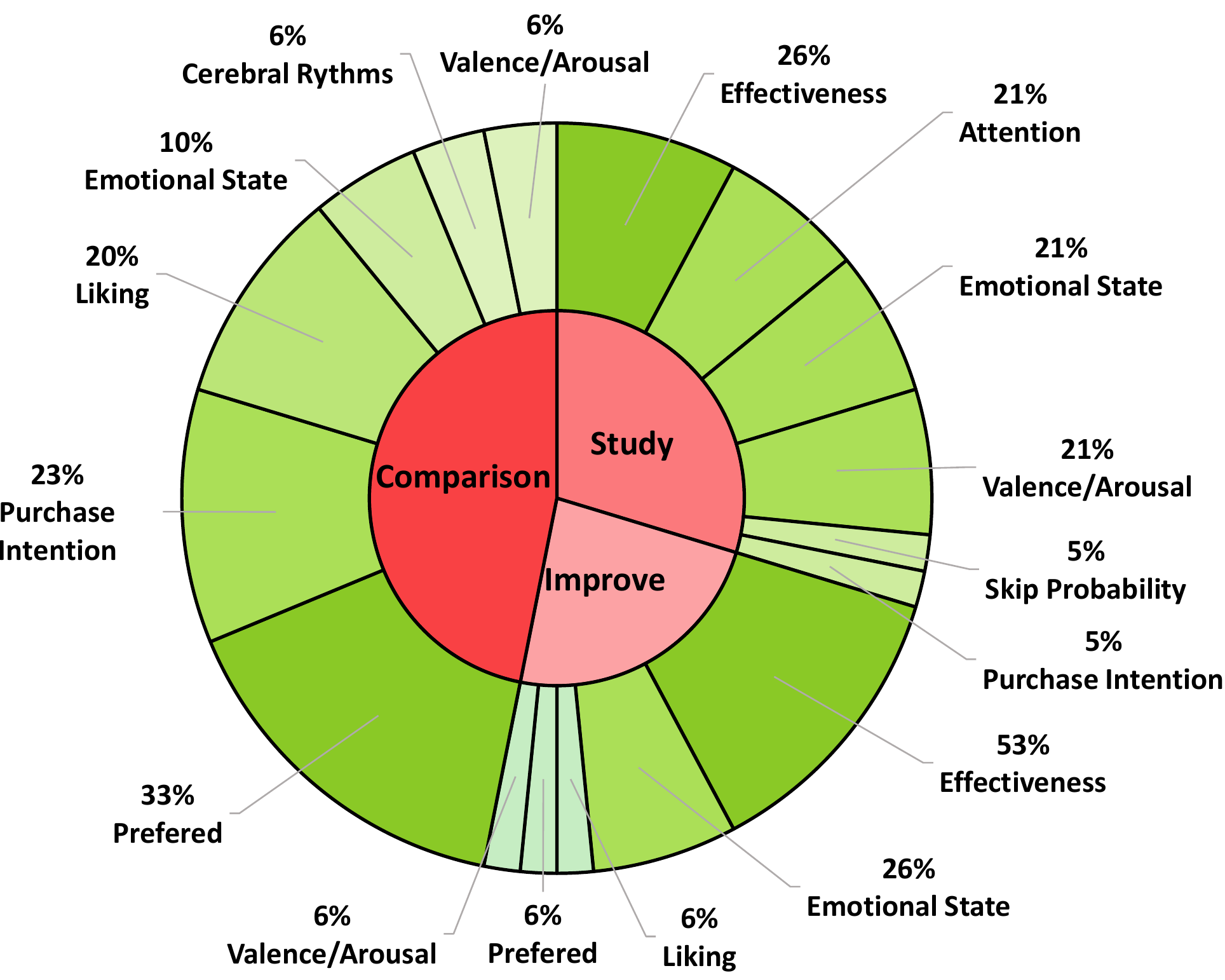}
     \caption{Metrics.}
     \label{fig:disobjetives}
    \end{subfigure}
    ~
    \begin{subfigure}{0.49\textwidth}
     \centering
     \includegraphics[width=\columnwidth]{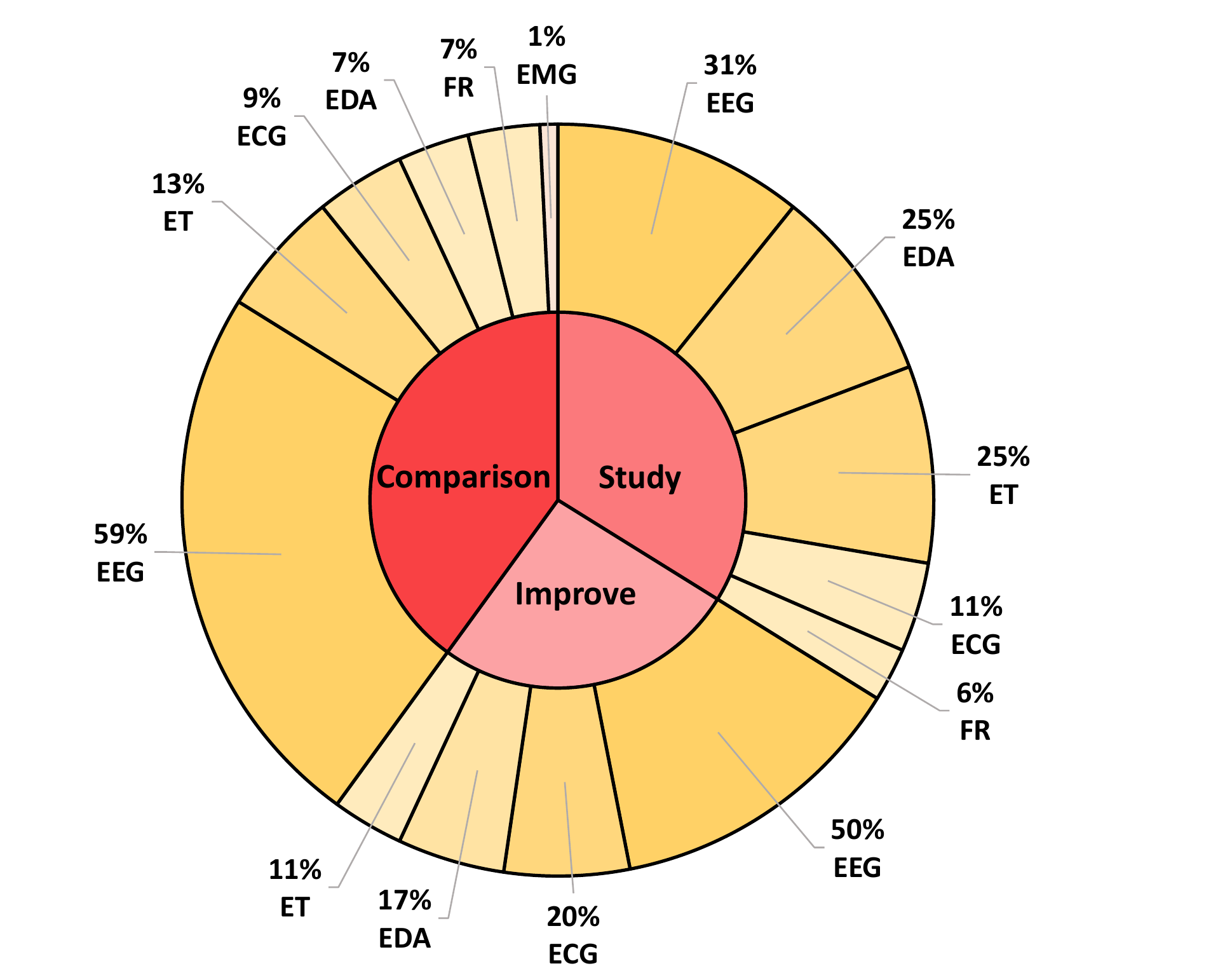}
     \caption{Devices and biosignals.}
     \label{fig:disbiosignals}
    \end{subfigure}
    \begin{subfigure}{0.49\textwidth}
     \centering
     \includegraphics[width=\columnwidth]{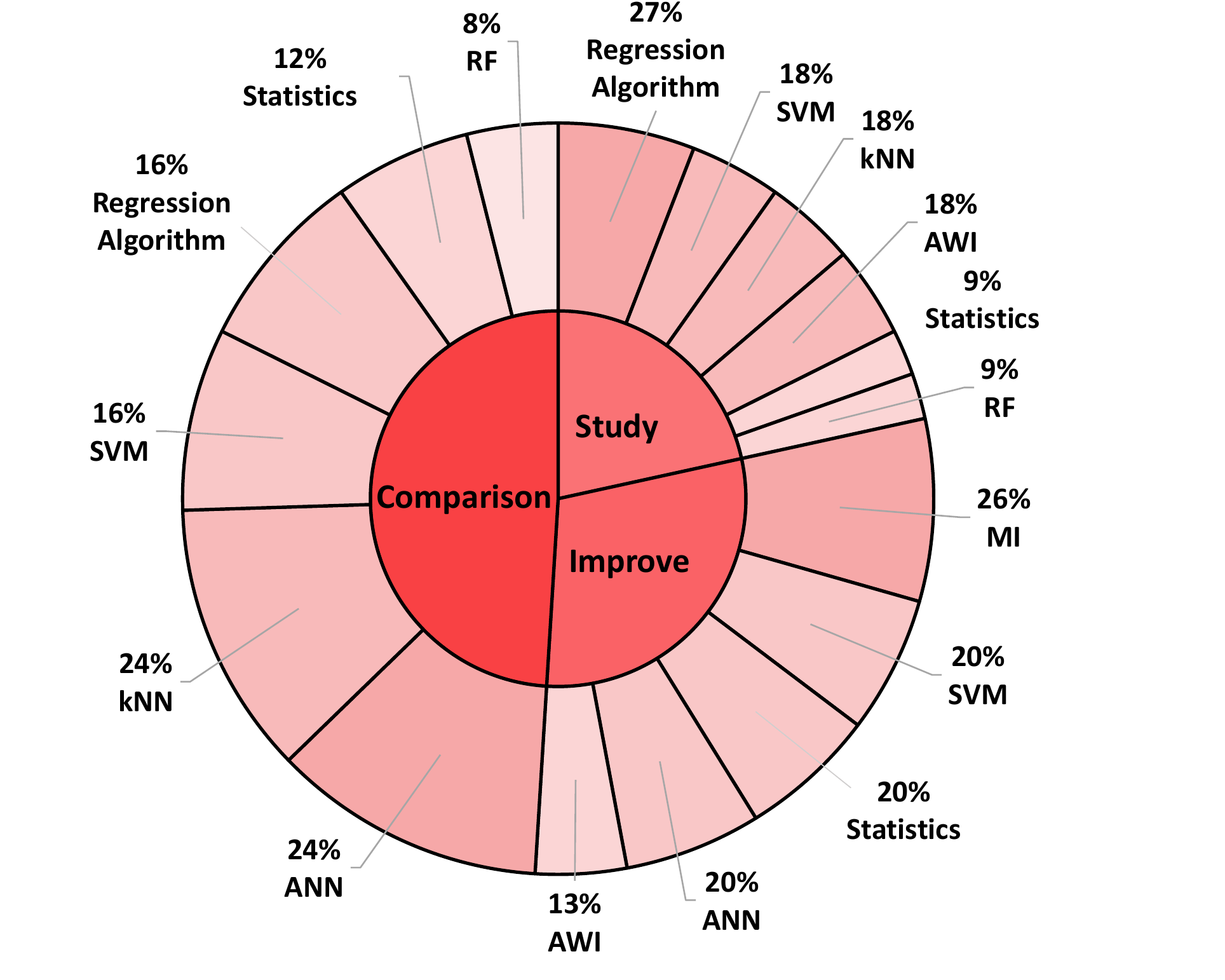}
     \caption{Automatic tools for hypothesis validation.}
     \label{fig:disml}
    \end{subfigure}
  
  \caption{Distribution of techniques and approaches for each objective.}
  \label{fig:relation}
\end{figure*}

\subsection{Objective of study}
\label{objectives}

The objective of a marketing campaign is fundamental when designing a campaign study. The literature mainly proposes three different objectives \cite{kumar2015neuromarketing, dragolea2011neuromarketing}. 

\begin{itemize}
    \item The first objective studies the impact of a given stimulus on the subject. Data is collected from the different biosensors to measure this impact, applying specific preprocessing and feature extraction techniques to obtain the relevant information. Once this information is available, classification algorithms can be used to measure the impact generated by the stimulus \cite{garczarek2021comparative}.
    
    \item The second objective focuses on improving the impact of the stimuli presented to users. In this approach, the same stimulus is usually presented with slight modifications while collecting the biosignal data to study how the modifications affect the stimuli and, thus, conclude whether the impact is improved \cite{alsmadi2021neuromarketing}.
    
    \item The third objective compares different advertisements, usually from the same market niche. This objective is the most common since it directly compares advertisements from different brands, obtaining which advertisements generate a more significant impact or a higher purchase probability \cite{caratu2020can}.
\end{itemize}

\figurename~\ref{fig:lifecycle} shows the distribution of the most used techniques in the different phases of the neuromarketing cycle. Specifically, for the distribution of the objectives shown in the first box, this figure highlights that the most proposed objective is the comparison between different advertisements. After this, the objectives focused on improving stimuli and assessing if their impact are distributed homogeneously.

\subsection{Hypothesis and metrics}
\label{hypothesis}

After presenting the objective of the study, it is necessary to define the hypothesis and metrics to measure whether or not the objective has been met. One of these mental states is the impact a given advertisement has on the subject, quantifying it as low, medium or high \cite{santos2020use}. The impact is one of the main objectives since a direct result can be obtained from the designed stimulus. Another objective is the purchase intention, or preference for a product \cite{huseynov2019incorporating, hakim2021machines}. These metrics are directly related to the sample objective of multiple similar products. In this way, the literature intended to recognize which advertisements generated a higher purchase probability. 

Other metrics are based on non-quantifiable human reactions, such as the emotional state they evoke or liking \cite{neomaniova2018dissonance, boksem2015brain}. The objective of detecting these states is aimed at studies about emotional neuromarketing, which seeks to represent past situations in order to increase the probability of purchase \cite{consoli2010new}. Usually, these states are measured by the values of valence and arousal values. For this reason, some studies attempt to predict these values to relate them to a particular mental state \cite{hernandez2017emotion, ogino2018mobile}. In this way, if a specific state of mind can be detected, it is possible to check whether the target has been designed correctly to achieve the proposed objective. Another point of interest studied in the literature is whether the subject intends to skip an advertisement (e.g., a web page) or not, as well as the attention that the subject is paying to a given advertisement \cite{libert2019predicting}.

As shown in the second box in the \figurename~\ref{fig:lifecycle}, the most studied metrics are the impact generated by an advertisement and the intention to purchase a product. This is mainly because these are the most interesting parameters for promoter companies. Other targets, such as liking or emotional state (directly related to brain bands), are of some interest in designing a stimulus. However, the main metrics differ depending on the objective for which they are proposed (see \figurename~\ref{fig:disobjetives}). As can be seen, for the study and improvement objectives, the main metric is to measure the effectiveness of a given advertisement, while for comparison, predicting which product is the favorite in a larger group is critical. Other elements of interest, such as attention generated, emotional state, or purchase intent, appear in the second position.
 
\subsection{Stimuli}
\label{stimulus}

The stimuli used in the related work is another crucial factor when designing a neuromarketing study. Because of this, this section reviews the different inputs used in the literature \cite{sebastian2014neuromarketing, fortunato2014review}. The first category corresponds to visual stimuli, which is based on continuously displaying a series of images to the user during a specific period \cite{de2020colour}. These images are usually commercial products to generate a reaction in the user that can be captured. The main advantage of this type of inputs is that the exposure time should not be very high, being also the most abundant type of stimuli in society, from billboards to social networks. However, the disadvantage of visual inputs is that they only provoke stimulation in the sense of sight so they may have less impact than others involving more senses. The second type of stimuli is audiovisual \cite{singh2020impact}. In particular, different sounds are added to the previously mentioned visual stimuli, which stimulate the sense of hearing, used commonly in TV. Thus, it can significantly impact the subject's attention since the load required to analyze the different inputs coming from various dimensions is more remarkable than if only one of them were involved. The disadvantage, however, is that the exposure time must be prolonged, which increases the probability of a loss of interest from the user.

In another direction, several works seek to stimulate senses other than the classic ones mentioned above, where some of these works use flavors to obtain a reaction from the user \cite{caratu2018application}. This type of stimulus is an excellent indicator to obtain consumers' reactions against food products, although it only applies to some real use cases. In another type of work, subjects interact with the product being publicized and, thus, get a response more aligned to reality \cite{bettiga2020consumers}.

Finally, due to the increase in online shopping in recent years, the literature has moved towards obtaining information about the design of a web page \cite{ungureanu2017neuromarketing}. Only visual stimuli can be used in these environments, so the colors, shapes, and distribution of the elements that compose it must be varied. Similarly to this trend, and thanks to the development of social networks, many companies have chosen to promote their products through online sites, doing it by ``influencers''. Influencers are people with the ability to influence potential buyers of a product or service by promoting or recommending the items on social media. Due to this, some studies seek to predict the impact of these promotional images or videos to advertise products \cite{manas2020neuromarketing}.

As seen in \figurename~\ref{fig:lifecycle}, the most used stimuli in the literature are visual and audiovisual. This usage is mainly because traditional advertising has been transmitted through television commercials or posters. Since most advertising has been done through the Internet recently, studies related to these techniques have gained popularity in recent years. \figurename~\ref{fig:disstimuli} relates the first phase of objectives, with the second phase of stimuli, showing relevant information on usability when selecting the best techniques.

\subsection{Devices and biosignals}
\label{devices}
Next, this work analyzes the devices and biosignals used in the studied literature \cite{bercea2012anatomy}. In particular, the authors employ diverse biosensors, such as BCIs, electrodermal sensors, cardiac event monitors, ET glasses, facial recognition cameras, and needle electrodes. 
 
 \begin{itemize}
     \item BCIs are systems that monitor or stimulate neural areas through electrodes. Concerning monitoring, the EEG signal \cite{kirschstein2009source} is collected based on the electrical exchanges produced by the neurons \cite{van2009brain}.
     
     \item Electrodermal sensors detect the changes in electrical (ionic) activity resulting from changes in sweat gland activity. The electrodes must be sensitive to these changes and able to transmit that information to the recording device. Specifically, this signal is known as EDA \cite{burstein1967primary}.
     
     \item Devices for measuring cardiac events have traditionally been used in the medical sector for measuring cardiovascular problems. It is possible to measure the heartbeat through heart rate measurement devices or obtaining an electrocardiogram (ECG) \cite{sameni2010review}.
     
     \item ET, meanwhile, use the reflection of the eyes in near-infrared light beams, thus detecting presence, attention, and focus, as well as the position of a person's eye and the pupil size \cite{dos2015eye}.
     
     \item Cameras for Facial Recognition (FR) allow recognition of a user's facial expressions in response to a stimulus \cite{espinoza2018neuromarketing}.
     
     \item Electromyography (EMG) uses electrodes to measure facial gestures. This signal acquires the electrical activity generated by the muscles \cite{balconi2011emotional}.
 \end{itemize}

Each of these biosignals can be employed for neuromarketing objectives thanks to the advantages they provide. In particular, the EEG signal is related to different mental states, such as emotions, attention, or stress \cite{subha2010eeg}. Through this relationship, the impact generated by an advertisement is known, which is beneficial in neuromarketing. Specifically, the EEG signal has several advantages, where one of the most relevant is that users cannot intentionally manipulate it, allowing the recognition of many mental states. However, the EEG signal has two major problems: on the one hand, it offers poor spatial resolution. Due to this, the EEG signal often encounters difficulties determining the exact source generating brain activity. Particularly, each electrode is responsible for covering a large area of neurons, being especially problematic when activity originates in different but nearby regions. On the other hand, the SNR or signal to noise ratio is low. It is a very low amplitude signal measured after passing through tissue, bone and scalp, and therefore very susceptible to be contaminated by artifacts. Therefore the preprocessing steps are very important to clean up the signal.
 
The EDA signal can measure the stress level of a user utilizing skin sweating. This reaction occurs due to stressful situations or other emotional states, widely employed in neuromarketing to study how users react to a stimulus \cite{goshvarpour2017accurate}. The advantage of these devices is that the users cannot consciously modify their reactions and are generally portable and comfortable. However, the main disadvantage of EDA is that it does not provide information about emotional valence and cannot determine whether the emotion is positive or negative. Since it only provides information on arousal, combining it with other techniques, such as EEG or ECG, is highly advisable. More precisely, the ECG signal can predict situations of stress or excitement in which the subject increases the state of alertness and heart pumping rate. The main advantage of this signal is that it has a direct relationship to stressful situations and, similarly to the devices used for EDA, it is portable. However, the ECG signal does not have enough value to represent a specific emotion, so it must be combined with other techniques.
 
In contrast, the ET signal can be related to states of interest through the time of fixation on an object. Examples of these states could be the attention, measured by the number of blinks; or emotional interest, measured by pupil dilation. Since these aspects are essential in neuromarketing studies, this signal is widely used in this area of knowledge \cite{mansor2018impact}. One of the main advantages of this signal is that it is adaptable to multiple environments. ET can also be used in passive tasks, such as viewing a television spot, and active tasks like evaluating the position of products on the shopping shelf while walking through the supermarket. Moreover, it can be combined with other devices since it is a sight-centered system and does not compete in space with EEG, EDA, or ECG. The main disadvantage of this signal is the need to combine it with other neuromarketing techniques since ET does not provide information on emotional valence and presents ambiguity when determining specific cognitive processes.
 
Finally, Facial Recognition (FR) is particularly suitable for detecting emotional valence, indicating whether the emotion is positive or negative. Although the FR signal has been used in various fields such as politics or the study of the facial expressions of listeners to a speech, it also has great importance in neuromarketing. This method has been widely used since it presents several advantages, such as a high speed where it provides real-time information on emotions, and a lack of intrusiveness as users do not need to wear a device. In addition, they are suitable for measuring the valence of mental states as it is a reliable indicator when detecting whether the stimulus produces positive or negative emotional reactions. However, as disadvantages, they are not a good option for measuring arousal or the intensity of emotions and they are very sensitive to users' movements. To overcome this, in some cases, the EMG signal is used. EMG is handy for recording first reactions to stimuli, which provides detailed information about the emotions produced \cite{levrini2021influence}. It can also help to understand how emotions are masked based on the differences in activity between voluntary and involuntary muscles. For this purpose, it can measure involuntary movements more accurately than facial gesture recognition. However, they are not commonly used due to their sensitivity to noise, and the required electrodes are invasive on the human body.

\figurename~\ref{fig:lifecycle} shows, in yellow, the usage distribution of each biosignal described above. It can be seen that EEG is the most widely used since the study of the signal has become popularized in recent years and is gaining momentum. Furthermore, the relationship of the different biosignals for each previously defined objective is shown in \figurename~\ref{fig:disbiosignals}. In the case of biosignals, a homogeneous distribution is observed, where the most used signal for all objectives is EEG. The second signal in popularity is the ET signal, except for the objective consisting in improving an advertisement, where the ECG signal is found.

\subsection{Data preprocessing techniques}
\label{preprocessing}

Once data sources are known and the previously explained sensors monitor the biosignals, these data are processed. This processing is necessary to eliminate noise signals produced by users' actions, such as flickering or involuntary movements, and to eliminate noise from external elements, such as the electrical network. The first of these techniques is the Bandpass filter \cite{ma2013low}, widely used in EEG signals, although it is also used for ECG and GSR \cite{xu2004high}. This filter allows defining a range of frequency bands and attenuating all those outside this range. Similarly, it can obtain only the target bands typically used for extracting the different brain rhythms. This filter can also be applied unilaterally, being able to attenuate only the upper bands, known as Lowpass filter \cite{gangkofner2007optimizing}, or the lower bands, named as Highpass filter \cite{karki2000active}. Another filter commonly used in the literature is the Notch filter, which allows the elimination of a given frequency range \cite{wang2014fully}. This way, it is possible to eliminate the frequency coming from the electrical network. There is also the possibility of using other types of filters more related to EEG, such as the Savitzky-Golay or Surface Laplacian filter \cite{acharya2016application, mcfarland2015advantages}. Employing these filters, it is possible to obtain a smoothed signal by calculating new data points.

In another way, specific channels of the EEG signal may give erroneous measurements for which different algorithms can be applied. In particular, EEG channel interpolation is of great significance when the EEG signal from a channel has low quality or is missing \cite{courellis2016eeg}. Furthermore, algorithms like Artifact Subspace Reconstruction (ASR) are commonly used to detect those channels providing erroneous data \cite{mullen2013real}. Typically, to deal with this problem of data density in EEG signals, the Downsampling algorithm is applied. This way, a series of less representative data can be obtained, which can be analyzed more efficiently, thus improving the results in terms of time in experiments seeking real-time processing \cite{paris2009fast}. A similar goal is pursued by the Common Average Reference (CAR) algorithm, which is a computationally simple technique suitable for on-chip and real-time applications. This algorithm is commonly used in EEG, where it is necessary to identify small signal sources in very noisy recordings \cite{ludwig2009using}. 

Other preprocessing algorithms, such as ICA, are typically used to detect individual components \cite{penny2000hidden}. In this direction, it is possible to eliminate components that interfere with the primary signal, such as flicker, and select those components to be kept for further classification. Similar to ICA, the ASR algorithm allows for an online and real-time evaluation, being a component-based method that can effectively remove transient or amplitude artifacts \cite{chang2018evaluation}.

Box five of \figurename~\ref{fig:lifecycle} shows the usage distribution of the different techniques among the papers studied. In most cases, Bandpass and Notch filters predominate because they are fundamental techniques applied to any biosignal. Moreover, techniques such as ICA, ASR, or Downsampling are algorithms centered on the EEG signal, so due to the large number of works that employ this signal, these techniques are widely used.

\subsection{Features extraction techniques}
\label{features}

Once the data have been processed, the relevant information is obtained from the data depending on the objective of the study. This information may be based on signal power, frequency, temporality, or a combination of them. For the case of biosignals like ECG, EDA, and EEG, the most commonly used techniques are those based on signal frequencies. For example, Fast-Fourier Transform (FFT) \cite{nussbaumer1981fast} is an efficient algorithm that allows calculating the Discrete Fourier transform (DFT) and the inverse. Other algorithms, such as Power Spectral Density (PSD) \cite{elson1995calculation}, allow studying how the power is distributed in the different frequencies of the signal. Another frequency-based algorithm is Discrete Wavelet Transform (DWT) \cite{shensa1992discrete}, which is a transform that decomposes a given signal into many sets, where each set is a time series of coefficients describing the time evolution of the signal in the corresponding frequency band. For a temporal analysis, the Detrended Fluctuation Analysis (DFA) feature is a method for determining the statistical self-affinity of a signal \cite{marton2014detrended}, useful for analyzing time series that appear to be long-term processes.

Other algorithms focus on obtaining certain features from the EEG signal. Among these, hemispheric symmetry \cite{venkatraman2015predicting} can obtain the Frontal Band Powers (FBP) \cite{yadava2017analysis} for each of the five frequency bands (Delta, Theta, Alpha, Beta, and Gamma) using the voltage difference between the front-lateral electrodes. Another feature that follows this example is the Global Field Power (GFP), which corresponds to the spatial standard deviation, and quantifies the amount of activity at each time point in the potential field. This feature considers the data from all recording electrodes simultaneously, resulting in a reference\-/independent descriptor of the potential field \cite{skrandies1990global}. An additional EEG-specific feature is the Individual Alpha Frequency (IAF), a promising electrophysiological marker of inter-individual differences in cognitive function. IAF has been related to trait-like differences in information processing and general intelligence and provides an empirical basis for the definition of individualized frequency bands \cite{corcoran2018toward}. Finally, algorithms such as mutual information analysis make it possible to detect changes in information transmission associated with change in sleep stages, and to understand how age affects the interdependence value. For other sensors such as functional Magnetic Resonance Imaging (fMRI), Inter-subject Correlations (ISC) \cite{kauppi2010inter} is a model-free approach to examine highly complex  data acquired in natural, context-dependent settings from audiovisual stimuli, such as during movie viewing.

For signals such as those from ET, it is necessary to consider other aspects of the signal. One of the most representative metrics is the Area of Interest (AOI) \cite{salvucci2000identifying}, which is a tool for selecting regions of a displayed stimulus and extracting metrics specifically for those regions. Although AOI is not strictly a metric, it defines the area by which other metrics are calculated. It is also possible to use heat maps to identify more visualized areas. These maps show the areas of an image where the user has paid more and less attention. Additional metrics are the characteristics of the images, such as intensity, contrast, saturation, or sharpness \cite{hernandez2017emotion}. These images are used to stimulate the users, and using this feature, it is possible to study if specific patterns in the image cause the same response in the users.

For simple signals such as EDA or ECG, algorithms are usually used to extract information from the amplitude and frequency of these signals. In the case of EDA, the most commonly used features are Skin Conductance Level (SCL) and Skin Conductance Response (SCR) \cite{braithwaite2013guide}. High SCL values are related to greater receptivity on the part of the subjects. In contrast, the SCR feature is very sparse and occurs when a relevant stimulus has been elicited. For the ECG signal, it is common to obtain features such as the heart rate, which represents the number of beats per second, as well as its variability \cite{nussinovitch2011reliability}.

Block six of \figurename~\ref{fig:lifecycle} shows the distribution of the use of these techniques for the extraction of characteristics, the most used being those related to power and frequency. This is mainly because they are common to most sensors, so their use is higher. Since ET is another of the most widespread sensors, techniques based on images and the position of the user's view are among the most common. Finally, valence estimation from user interaction features is the least used, as they can be very susceptible to the user's subjectivity.

\subsection{Automatic tools for hypothesis validation}
\label{classification}

Once the relevant data information is available, it is necessary to classify them to define to which state they are associated with. The works on neuromarketing follow two different trends for data classification: extracting statistics from the data to be studied by experts and applying ML/DL algorithms. First, researchers obtain statistical values and represent the data graphically, allowing the identification of the different mental states visually. This situation has been widely studied in scenarios with ECG, GSR, and EEG biosignals where specific statistical values such as mean, standard deviation, or median, among others, can be directly related to an intention or state. Another of these values can be the Memorization Index (MI) \cite{davidson2004does}, which is obtained by calculating the mean of the values of the theta band, which can be directly translated into this index. In the same way, the Approach-Withdrawal Index (AWI) \cite{maglione2015alpha} defines the asymmetry of Pre-Frontal Cortex Activity that is closely linked to the feeling of pleasantness experienced by the subject during sensorial stimulation.

In the second case, where the authors apply ML algorithms, the most common are Support Vector Machines (SVM) \cite{noble2006support}, k-Nearest Neighbors (kNN) \cite{peterson2009k}, and Random Forest (RF) \cite{fawagreh2014random}. These algorithms perform relatively well in most scenarios, with low computational consumption, also achieving high accuracy values with biological data types such as EEG \cite{dey2018machine}. 

Another type of algorithm that belongs to a subset of AI, is DL algorithms. These algorithms based on neural networks have become more relevant in recent years. This usage increase is mainly because of the advance of biosensors, and the large amounts of data handled in neuromarketing studies. In addition, the time factor for evaluation and the lack of need for feature extraction make this type of AI algorithm a candidate for solving the time problem with feature extraction. Because of these advantages, these algorithms are of great value in scenarios where a large amount of data is received, and evaluation times must be reduced \cite{ganapathy2018deep}. This is the case of Convolutional Neural Networks (CNNs) \cite{yamashita2018convolutional}, Probabilistic Neural Network (PNN) \cite{specht1990probabilistic} and Artificial Neural Networks (ANNs) \cite{krogh2008artificial}.

As can be seen in block seven of \figurename~\ref{fig:lifecycle}, the the most popular algorithms in the literature are SVM, kNN, and RF, followed by statistics studies and Regression Algorithms (RA). Finally, DL-based algorithms gained popularity, although their use is not too elevated due to their restrictions because of the large amount of data they need. The use of these algorithms for each objective is illustrated in \figurename~\ref{fig:disml}. Artificial Neural Networks (ANN) have also been extensively applied for advertisement matching, followed by traditional ML algorithms such as kNN and SVM. Conversely, the most common techniques for advertisement enhancement objectives are based on Memorization Index (MI) and visual statistics study. Regression algorithms are the most used for the study objective since they predict particular numerical values of an advertisement, such as a rating between zero and ten. In general, the preferred algorithms are kNN and SVM, which are traditional ranking algorithms.

\section{What combinations of data from biosignals, stimuli, and processing and classification techniques are suitable for each objective?}
\label{comparison}
This section compares the collected articles divided by the neuromarketing objectives for which they have been designed. Section \ref{objectives} already defined three different neuromarketing objectives: (i) study of the impact of an advertisement, (ii) improve an advertisement, and (iii) comparison between advertisements. With these three points, it is possible to study the works with a common purpose and compare the techniques used within them and the results they have obtained.

\subsection{Study of the impact of an advertisement}

This objective is based on collecting different biosignals representing the user's mental states and thus relating it to a lower or higher impact. \tablename~\ref{tab:study} shows the works collected, grouped by the stimuli used, also highlighting aspects of the papers taken into account, such as the number of subjects participating in the experimentation, the type of stimulus used, the preprocessing techniques, feature extraction, and classification techniques. Finally, the number of states recognized in the classification and the results gathered by each study are considered.

In order to simulate the most common stimuli encountered in real life, some researchers employ audiovisual stimuli. The work performed by Dimpfel et al., \cite{dimpfel2015neuromarketing} aimed to determine whether EEG biosignals and ET are effective methods for neuromarketing studies. The FFT algorithm was applied in the case of the EEG signal, and heat map in the case of ET. The classification was done visually by the researchers, obtaining statistics from them and concluding that these algorithms are suitable for this purpose. Later, Cartocci et al., \cite{cartocci2017electroencephalographic} carried out a study combining EEG with ECG and EDA sensors, using IAF and GFP characteristics to obtain values related to the frequencies in the data. For the analysis of the results, the values produced by the AWI algorithm were obtained, relating the results to the effectiveness or ineffectiveness of the different advertisements. 

Other papers focus on using audiovisual stimuli and seek to study the impact of specific parameters on users, such as colors or the positioning of objects. The first work was performed by Cosic et al., \cite{cosic2016neuromarketing}, focusing on obtaining the best position for placing objects and studying which is the best composition of objects on the screen. For this research, an ET biosensor was applied to 62 subjects. Statistics were extracted from the data produced by AOI, concluding that the best position is the center. In the same direction, Baraybar et al., \cite{baraybar2017evaluation} and Hamelin et al., \cite{hamelin2017emotion} focused on emotional aspects that the advertisements evoke in the subjects, using ECG, EDA, and FR since they directly relate to the different emotional states. These works used the first approach to classification based on collection of statistics and visualization by the researchers. The results obtained by Hamelin et al., \cite{hamelin2017emotion} show that the advertisements that evoke emotions are the ones that obtain the best results. Specifically, Baraybar et al., \cite{baraybar2017evaluation} concluded that the emotion obtaining the best results is sadness.

Other authors aim to detect more general aspects of an advertisement with audiovisual stimuli, such as the effectiveness of an advertisement, or the likelihood that a given advertisement will be skipped, as previously presented in Section \ref{hypothesis}. Concerning effectiveness, Cartocci et al., \cite{cartocci2016pilot} conducted a study on how audiovisual stimuli affect neural response in subjects. For this, they used a relatively low number of seven subjects and obtained their EEG signals. This signal for each subject was treated with the Notch filter and the ICA algorithm, once the external noise has been removed, the relevant information is gathered from it using the IAF and GFP algorithms. They classified this information by using the AWI algorithm as effective or ineffective. The results show that the activation generated in each of the hemispheres can be related to the effectiveness of the advertisement on the subject. In the case of purchasing intention, Sung et al., \cite{sung2021opening} used EEG, EDA, ECG, and ET signals obtained from 142 users. For noise removal, the authors applied the Notch and Bandpass filters, after which they extracted the relevant information using the FFT algorithm from the EEG, EDA, and ECG signals. They concluded that consumers’ experience of anger underlies their negative responses toward the brand when it was marketed to be luxurious. Finally, Libert et al., \cite{libert2019predicting} gathered entropy and CV2 values from the EEG signal acquired from four users to use a series of ML algorithms, such as RF, SVM, and kNN. The objective was to predict the probability of skipping a given advertisement, obtaining 75.8\% accuracy. 

Other research focuses on predicting numerical values when assessing the quality of a stimulus in neuromarketing studies from audiovisual stimuli, such as a value that represents how much a user liked the advertisement. Following this objective, Ogino et al., \cite{ogino2018mobile} sought to predict the valence from the EEG signal and the different bands obtained by FFT. They collected this signal from 30 subjects, obtaining a 72.40\% average accuracy using LR, kNN, SVM, and SVR algorithms. Another numerical value that can be predicted is arousal, where Michael et al., \cite{michael2019study} investigated the prediction of this value using EEG signals and ET. These data were collected from 30 participants, obtaining statistics about the cognitive load generated and visually estimating the arousal value generated. In addition, Hernandez et al., \cite{hernandez2017emotion} aimed to predict when the arousal value was low or high employing the EDA biosignal value. This study used public datasets (DEAP) and preprocessed them with the Bandpass filter, Notch filter, and the ICA algorithm. After this, this work extracted features from the displayed images, applying a regression algorithm to predict a specific value of the EDA signal.

Another category of stimuli applicable to study objective is visual stimuli. Similar to the previously studied works, some authors have focused on studying the characteristics of images. In pursuit of this goal, Cuesta et al., \cite{cuesta2018neuromarketing} studied the best color for generating a positive reaction by the subject, using for that ET, EDA, and FR signals. Data such as heat maps, fixation points, or facial expressions were gathered from these signals. After this, the authors conducted a visual study of this information on five subjects, concluding that the color that attracted the subjects' attention the most was yellow. Similarly, Santos et al., \cite{santos2020use} focuses on the analysis of modification a logo, such as the logo of the brand ``Worten''. They used four different biosignals for this improvement: EEG, EDA, ECG, and ET. From these signals, they obtained frequency-based data from EEG, EDA, and ECG signals using the PSD algorithm. In addition, they extracted information from the EEG signal by applying the AWI algorithm and a heat map using ET. All this information was studied visually in order to know the impact of the logo. Furthermore, an area of interest is the analysis of subjects' reactions to images utilizing EEG, EDA, and ET, such as the one carried out by Pop et al., \cite{pop2013using} or Bastiaansen et al., \cite{bastiaansen2018my}. The former collected these data from a considerably large number of subjects, 44. They extracted the relevance from these data to measure the subjects' reactions. In the second paper, the researchers gathered the EEG signal from 32 subjects, extracted valence and arousal features, and classified the results visually. Another work focused on predicting how much users liked an advertisement is the work done by Kumar et al., \cite{kumar2019fusion}. In this work, the aim is to predict a value between zero and five using an RF regression algorithm.

Following the growing demand for online shopping in recent years, articles related to the study of web page designs have emerged. Kvasnicova et al., \cite{kvasnicova2016investigation} aimed to show if EEG and ET signals were valid for this study. For this, they collected data on ten subjects and applied algorithms to study frequencies such as PSD and STFT in the case of the EEG signal. Additionally, they obtained AOI data from the ET signal. These data were analyzed visually, concluding that these techniques are helpful. In the same direction, Ungureanu et al., \cite{ungureanu2017neuromarketing} replaced the EEG signal with the EDA signal. In this case, special attention was paid to the ET biosignal by obtaining data such as AOI and pupil size. This work corroborates again that these methods are valid. Similarly, Slanzi et al., \cite{slanzi2017combining} sought to predict which areas of a webpage were most likely to receive clicks. This prediction was based on data from the webpage's saliency maps combined with the subjects' EEG signal data. The final accuracy obtained was over 70\%.

Finally, some authors have tried to explore new types of stimuli. Cuesta et al., \cite{cuesta2018case} proposed a series of experiments to study subjects' reactions to music. EDA, ET, and FR were used, together with a heat map and AOI. The results show that music increases the impact of a product on the subjects. Another stimulus is based on generating a cognitive load in the subjects by performing different tasks, such as choosing a particular vending machine. Sargent et al., \cite{sargent2020neuroergonomic} explored the preference for a particular brand when selecting a vending machine to purchase coffee. For this purpose, EEG and EDA biosignals were used, together with PSD. For the classification of the results, frequency-related statistics are obtained, subsequently performing a visual classification. The results demonstrate that the machines whose brands are the market leaders or, more extended, obtain greater consumer acceptance.

\begin{table*}[!ht]
  \caption{Works that seek to study the impact of an advertisement.}
  \label{tab:study}
  \resizebox{\textwidth}{!}{
  \begin{tabular}{@{}cccccccccc@{}}
    \toprule
    \textbf{Ref} & \textbf{Year} &  \textbf{Biosignal} & \textbf{Subjects} &\textbf{Stimuli} & \textbf{Preprocessing} & \begin{tabular}[c]{@{}c@{}}\textbf{Feature} \\ \textbf{Extraction}\end{tabular} & \begin{tabular}[c]{@{}c@{}}\textbf{Classification} \end{tabular} & \begin{tabular}[c]{@{}c@{}}\textbf{Predicted} \\ \textbf{States}\end{tabular} & \textbf{Results}\\
    \midrule
    \hline
    
    \cite{dimpfel2015neuromarketing}& 2015 & \begin{tabular}[c]{@{}c@{}} EEG \\ ET \end{tabular} & 10 & Audiovisual & Not specified & \begin{tabular}[c]{@{}c@{}} FFT \\ Heat map \end{tabular} &  \begin{tabular}[c]{@{}c@{}} Value \\ display \end{tabular} &  Not specified & \begin{tabular}[c]{@{}c@{}} Neurocode-Tracking with ET\\ can be successfully used\\ in advertisement research and\\ gives valid information \end{tabular}\\
    \hline
    
    \cite{cartocci2016pilot}& 2016 & EEG &  7 & Audiovisual & \begin{tabular}[c]{@{}c@{}} Notch filter \\ ICA \end{tabular} & \begin{tabular}[c]{@{}c@{}} IAF \\ GFP \end{tabular} & AWI & \begin{tabular}[c]{@{}c@{}} Effective \\ Ineffective \end{tabular} & \begin{tabular}[c]{@{}c@{}} Increased deactivation \\ in the right dorsal \\ and  increased left frontal  \\gyrus activation \end{tabular} \\
    \hline
    
    \cite{cosic2016neuromarketing}& 2016 &  ET & 62  & Audiovisual & Not specified & AOI & Statistics & \begin{tabular}[c]{@{}c@{}}Focus \\ of attention \\ in screen\end{tabular} & \begin{tabular}[c]{@{}c@{}}Position of objects\\ in the center of the screen\\ obtains the greatest impact\end{tabular} \\
    \hline
    
    \cite{baraybar2017evaluation} & 2017 & \begin{tabular}[c]{@{}c@{}}ECG \\ EDA\end{tabular} & 36 & Audiovisual & Bandpass & No specefied & Recall & \begin{tabular}[c]{@{}c@{}}Emotional \\ State\end{tabular} & \begin{tabular}[c]{@{}c@{}}Advertisements that evoke\\ sadness have greater\\ impact\end{tabular} \\
    \hline
    
    \cite{cartocci2017electroencephalographic}& 2017 & \begin{tabular}[c]{@{}c@{}} EEG \\ ECG \\ EDA \end{tabular} &  22 & Audiovisual & \begin{tabular}[c]{@{}c@{}} Bandpass filter \\ Notch filter \\ ICA \end{tabular} & \begin{tabular}[c]{@{}c@{}} IAF \\ GFP \end{tabular} & AWI & \begin{tabular}[c]{@{}c@{}} Awarded \\ Effective \\ Ineffective \end{tabular} & \begin{tabular}[c]{@{}c@{}} Fear-based campaigns with\\ complex narrative stories reported\\ worse results than those with\\ a narrative o experiential communication style. \end{tabular}  \\
    \hline
    
    \cite{hamelin2017emotion} & 2017 & FR & 60 & Audiovisual & Not specified & \begin{tabular}[c]{@{}c@{}}High Valence \\ Low Valence\end{tabular} & Statistics & \begin{tabular}[c]{@{}c@{}}Attitude \\ score\end{tabular} & \begin{tabular}[c]{@{}c@{}}Advertisements designed to\\ evoke emotions have\\ greater impact\end{tabular}\\
    \hline

    \cite{hernandez2017emotion}& 2017 & EDA &  \begin{tabular}[c]{@{}c@{}} DEAP \\ MAHNOB-HCI  \end{tabular} & Audiovisual & \begin{tabular}[c]{@{}c@{}} Bandpass filter \\ Notch filter \\ ICA \end{tabular} & \begin{tabular}[c]{@{}c@{}} Image \\ Characteristics \end{tabular} & \begin{tabular}[c]{@{}c@{}} Regression \\ Algorithm \end{tabular} & \begin{tabular}[c]{@{}c@{}} High Attention \\ High Arousal \end{tabular} & \begin{tabular}[c]{@{}c@{}} Value \\ of EDA \end{tabular} \\
    \hline
    
    \cite{ogino2018mobile}& 2018 & EEG &  30 & Audiovisual & Bandpass filter & FFT & \begin{tabular}[c]{@{}c@{}} LR \\ kNN \\ SVM \\ SVR \end{tabular} & Valence & 72.40\% of accuracy \\
    \hline
    
    \cite{neomaniova2018dissonance}& 2018 & EEG &  16 & Audiovisual & Bandpass filter & FFT & Statistics & Emotional State & \begin{tabular}[c]{@{}c@{}} During the video watching  \\ dominated the neutral feelings, \\ while the results of the electrical brain \\ activity measurements indicated \\ prevailing negative emotions \end{tabular} \\
    \hline

    \cite{libert2019predicting}& 2019 & EEG &  4 & Audiovisual & \begin{tabular}[c]{@{}c@{}} Bandpass filter \\ Downsampling \end{tabular} & \begin{tabular}[c]{@{}c@{}} Entropy \\ Power-Based CV2 \end{tabular} & \begin{tabular}[c]{@{}c@{}} RF \\ kNN \\ SVM \end{tabular} & \begin{tabular}[c]{@{}c@{}} Skip \\ No Skip \end{tabular} & 75.8\% of accuracy \\
    \hline

    \cite{michael2019study}& 2019 & \begin{tabular}[c]{@{}c@{}} EEG \\ ET \end{tabular} &  30 & Audiovisual & Not specified & Cognitive load & Not specified & Arousal & \begin{tabular}[c]{@{}c@{}} Differences in ET and EEG\\ signals depending on the\\ tourist destination presented \end{tabular} \\
    \hline
    
    \cite{sung2021opening}& 2019 & \begin{tabular}[c]{@{}c@{}} EEG \\EDA \\ECG \\  ET \end{tabular} &  142 & Audiovisual & \begin{tabular}[c]{@{}c@{}} Bandpass filter \\ Notch filter \end{tabular} & FFT & Statistics & \begin{tabular}[c]{@{}c@{}} Purchase \\intention\end{tabular} & \begin{tabular}[c]{@{}c@{}} High arousal \\when seeing \\ non-luxury \\ brand stores\end{tabular} \\
    \hline
    
    \cite{pop2013using}& 2013 & \begin{tabular}[c]{@{}c@{}} EEG \\ EDA \\ ET \end{tabular} &  44 & Image & Not specified & Relevance & Not specified & Reaction & \begin{tabular}[c]{@{}c@{}}  The honey packaging generates \\  a positive reaction but \\  it needs a long time to do that \end{tabular} \\
    \hline
    
    \cite{cuesta2018neuromarketing}& 2018 & \begin{tabular}[c]{@{}c@{}}EDA \\ ET \\ FR \end{tabular} & 5 & Image & Not specified & \begin{tabular}[c]{@{}c@{}} Heat map \\ Total time spent \\ Peaks of EDA \\ Facial expressions \end{tabular} &  \begin{tabular}[c]{@{}c@{}}Visual \\ study\end{tabular} & \begin{tabular}[c]{@{}c@{}}Colors that generate\\ the greatest impact\end{tabular} & \begin{tabular}[c]{@{}c@{}}Yellow is the color\\ that generates the\\ greatest impact\end{tabular} \\
    \hline

    \cite{kumar2019fusion}& 2018 & \begin{tabular}[c]{@{}c@{}} EEG \end{tabular} &  26 & Image & \begin{tabular}[c]{@{}c@{}} Highpass filter \\ Lowpass filter \\ ASR \end{tabular} & DWT & RF Regression & Rating & \begin{tabular}[c]{@{}c@{}} Product liking value \end{tabular} \\
    \hline
    
    \cite{bastiaansen2018my}& 2018 & EEG &  32 & Image & \begin{tabular}[c]{@{}c@{}} Bandpass filter \\ Automatic Artifacts Removal \end{tabular} & \begin{tabular}[c]{@{}c@{}} Valence \\ Arousal \end{tabular} & Statistics & \begin{tabular}[c]{@{}c@{}} Study of reactions to images\\ after having been previously\\ influenced positively and negatively \end{tabular} & \begin{tabular}[c]{@{}c@{}} Subjects positively\\ influenced in a previous way\\ generate ERPs\end{tabular} \\
    \hline
   
    \cite{garcia2019isolated}& 2019 & \begin{tabular}[c]{@{}c@{}} EEG \\  ET \end{tabular} &  40 & Image & Not specified & \begin{tabular}[c]{@{}c@{}} Alpha-Band \\ Oscillation \\ AOI \end{tabular} & Not specified & \begin{tabular}[c]{@{}c@{}} Correlation \\ between attention \\ levels and preferences \end{tabular} & \begin{tabular}[c]{@{}c@{}} Consumers' eye movements could \\  not be employed as a \\ parameter to predict \\  packaging preferences \end{tabular} \\
    \hline
    
    \cite{santos2020use}& 2020 & \begin{tabular}[c]{@{}c@{}} EEG \\EDA \\ECG \\  ET \end{tabular} &  52 & Image & \begin{tabular}[c]{@{}c@{}} Bandpass filter \\ ICA \end{tabular} & \begin{tabular}[c]{@{}c@{}} PSD \\ AWI \end{tabular} & Not specified & \begin{tabular}[c]{@{}c@{}} Impact \\of logo \end{tabular} & \begin{tabular}[c]{@{}c@{}} Color improves user \\response to advertisements \end{tabular} \\
    \hline
    
    \cite{kvasnicova2016investigation}& 2016 & \begin{tabular}[c]{@{}c@{}} EEG \\ET \end{tabular} &  N.E & Web pages & Not specified & \begin{tabular}[c]{@{}c@{}} PSD \\ STFT \\ AOI \end{tabular} & \begin{tabular}[c]{@{}c@{}} Visual \\ study \end{tabular} & \begin{tabular}[c]{@{}c@{}} Study of \\ webpage \end{tabular} & \begin{tabular}[c]{@{}c@{}} Higher brain activity\\ only during decision\\ making processes\end{tabular} \\
    \hline

    \cite{ungureanu2017neuromarketing}& 2017 & \begin{tabular}[c]{@{}c@{}} EDA \\ ET \end{tabular} & 10 & Web Pages  & Not specified & \begin{tabular}[c]{@{}c@{}} Heat map \\ Pupil Size \end{tabular} & Not specified & \begin{tabular}[c]{@{}c@{}} AOI \\ Time spent \end{tabular}  & \begin{tabular}[c]{@{}c@{}} Human emotions and visual attention\\ are highly correlated with \\practical marketing applications \end{tabular}   \\
    \hline

    \cite{slanzi2017combining}& 2017 & \begin{tabular}[c]{@{}c@{}} EEG \\ ET \end{tabular} & 21 & Web Pages  & Not specified & \begin{tabular}[c]{@{}c@{}} Heat map \\ Frequency bands \end{tabular} & DWT & \begin{tabular}[c]{@{}c@{}} Saliency \end{tabular}  & \begin{tabular}[c]{@{}c@{}} 71.09\% accuracy \end{tabular}   \\
    \hline

    \cite{cuesta2018case}& 2018 & \begin{tabular}[c]{@{}c@{}} EDA \\ ET \\ FR \end{tabular} & 19 & Music & Not specified & \begin{tabular}[c]{@{}c@{}} heat map \\ AOI \end{tabular} & Not specified & \begin{tabular}[c]{@{}c@{}} Enjoy \\ Engagement \end{tabular} & \begin{tabular}[c]{@{}c@{}} Music \\ improves \\the impact \end{tabular}\\
    \hline
    
    \cite{sargent2020neuroergonomic}& 2020 & \begin{tabular}[c]{@{}c@{}} EEG \\EDA \end{tabular} &  26 & Cognitive task & \begin{tabular}[c]{@{}c@{}} Highpass filter \\ Lowpass filter \\ ASR \end{tabular} & PSD & Statistics & Valence & \begin{tabular}[c]{@{}c@{}} Preference for machines\\ containing market-leader brands\end{tabular} \\
    \hline

    \bottomrule
  \end{tabular}}
\end{table*}

\subsection{Improve the impact of an advertisement}

The second objective is based on improving an advertisement by modifying the stimuli. In contrast to the previous goal, where the main objective is only to study how a given stimulus affects users, in this one, different tests are performed varying these stimuli to improve the acceptance of the target audience in general. The works analyzed are grouped according to the same strategy followed in the previous section, depending on the type of stimulus used. These papers are listed in \tablename~\ref{tab:improve}, which summarizes the main aspects of each of them.

Starting with audiovisual stimuli, some papers focus on studying the subjects' reactions, classifying these reactions as positive or negative. Among these, the work by Colomer et al., \cite{colomer2016comparison}, used EEG, ECG, and EDA biosignals from 47 subjects, together with Global Field Power (GFP) and PSD techniques for feature extraction. After this, they applied ML algorithms obtaining up to 87.62\% accuracy to describe the effectiveness of a given advertisement. Similarly, Emsawas et al., \cite{emsawas2019feasible}, conducted a study using EEG, ECG, and ET, applying only PSD and ML algorithms, where they obtained 76.4\% accuracy.

Other authors seek to predict when an advertisement has generated user acceptance by measuring liking. In this way, it is possible to vary the stimuli to obtain the best combination and thus increase the impact of the advertisement. Mateusz and Kesra \cite{mateusz2020cognitive} proposed a study using EEG, ECG, and EDA sensors to test whether these techniques are applicable in this objective. They extracted information from the data using GFP and IAF algorithms, subsequently using MI, AW, and EI algorithms for classification. The authors concluded that these methods are acceptable in this research field. Complementary, Martínez et al., \cite{martinez2018measuring} employed EEG, ECG, EDA, and ET biosignals to study the visual fixation points for the recall of each of the products shown. For this, they applied the GFP algorithm to obtain the information from the different biosignals and the MI and EI algorithms for the classification.

Another relevant dimension for researchers is the study of specific aspects of the advertisements, such as the colors used. Lee et al., \cite{lee2014spell} studied how the green color can improve an advertisement and how it affects consumers through the EEG biosignal. For that, they applied FFT and GLM and, after obtaining statistics, they concluded that regular consumers of green products have a greater activation when exposed to this type of product. In contrast, Ohme et al., \cite{ohme2010application} performed a study on how to improve the potential of advertisements shown on television. This experiment was performed on 45 users, obtaining the EEG signal, preprocessed using the ICA algorithm, and extracting the relevant information employing the FFT algorithm. For the classification, a mean classifier was applied, indicating that certain advertisements generate an activation peak in the frontal lobe.

In the case of image stimuli, there is a similar trend to those created with audiovisual stimuli. For example, some solutions study the visual design of products, such as colors, patterns, or combinations of these characteristics. Khushaba et al., \cite{khushaba2012choice} completed an experiment for color modification using the EEG signal with GLM and a Mutual Information (MI) classifier. The results showed that the different combinations of these characteristics have a tangible impact on user perception. Moreover, Matukin et al., \cite{matukin2016toward} did a similar work defining the best techniques for the visual design of advertisements. For this, they used EEG and ET signals together with FFT.

Finally, Aldayel et al., \cite{aldayel2020deep} focused their research on predicting the subjects' reactions with EEG signals. The authors used PSD values and ML algorithms such as RF, kNN, or SVM for this. They obtained up to 94\% accuracy in predicting these reactions from users. Similarly, Samsuri et al., \cite{samsuri2016left} aimed to obtain the level of users' attention when viewing an advertisement using EEG and ET signals, thus studying the evoked potentials generated by these stimuli. Likewise, Golnar et al., \cite{golnar2019application} aimed to predict whether users liked or disliked an advertisement. For this, they used PSD values collected from the EEG signal. Regarding classification, they used SVM and LDA algorithms, obtaining up to 95\% accuracy for predicting these states.

\begin{table*}[!ht]
  \caption{Works that seek to improve an advertisement.}
  \label{tab:improve}
  \resizebox{\textwidth}{!}{
  \begin{tabular}{@{}cccccccccc@{}}
    \toprule
    \textbf{Ref} & \textbf{Year} &  \textbf{Biosignal} & \textbf{Subjects} &\textbf{Stimuli} & \textbf{Preprocessing} & \begin{tabular}[c]{@{}c@{}}\textbf{Feature} \\ \textbf{Extraction}\end{tabular} & \begin{tabular}[c]{@{}c@{}}\textbf{Classification} \end{tabular} & \begin{tabular}[c]{@{}c@{}}\textbf{Predicted} \\ \textbf{States}\end{tabular} & \textbf{Results}\\
    \midrule
    \hline
        
    \cite{ohme2010application}& 2010 & EEG & 45 & Audiovisual & ICA & FFT & Mean classifier  & \begin{tabular}[c]{@{}c@{}} Potential of TV \\ advertisements\end{tabular}  & \begin{tabular}[c]{@{}c@{}} Statistically significant difference \\ among the three advertisements  \end{tabular} \\
    \hline
    
    \cite{lee2014spell}& 2014 & EEG &  19 & Audiovisual & \begin{tabular}[c]{@{}c@{}} Bandpass filter \\ Notch filter \end{tabular} & \begin{tabular}[c]{@{}c@{}} FFT \\ GLM \end{tabular} & Statistics & \begin{tabular}[c]{@{}c@{}} Study the \\ impact of \\ the green color \end{tabular} & \begin{tabular}[c]{@{}c@{}} Frontal theta activations were \\ higher among eco-friendly subjects\end{tabular} \\
    \hline
    
    \cite{colomer2016comparison}& 2016 & \begin{tabular}[c]{@{}c@{}} EEG \\ ECG \\ EDA  \end{tabular} &  47 & Audiovisual & \begin{tabular}[c]{@{}c@{}} Bandpass filter \\  ICA \\ Adjust\end{tabular} & \begin{tabular}[c]{@{}c@{}} GFP \\ PSD \end{tabular} & \begin{tabular}[c]{@{}c@{}} MCC \\ BAG \\ RF \\ ASC \end{tabular} & \begin{tabular}[c]{@{}c@{}} Positive \\ Neutral \\ Negative \end{tabular} & 87.62\% of accuracy\\
    \hline
    
    \cite{martinez2018measuring}& 2018 & \begin{tabular}[c]{@{}c@{}} EEG \\ ECG \\ EDA \\ ET \end{tabular} & 22 & Audiovisual & \begin{tabular}[c]{@{}c@{}} Bandpass filter \\ Notch filter\\ ICA \end{tabular} & GFP & \begin{tabular}[c]{@{}c@{}} MI \\ EI \end{tabular} & \begin{tabular}[c]{@{}c@{}} Pleasant \\ Unpleasant \end{tabular} & \begin{tabular}[c]{@{}c@{}} The number of fixations affects \\  the recall of the showed products \end{tabular} \\
    \hline
    
    \cite{emsawas2019feasible}& 2019 & \begin{tabular}[c]{@{}c@{}} EEG \\ ECG \\ ET \end{tabular} &  130 & Audiovisual & Not specified & PSD & \begin{tabular}[c]{@{}c@{}} LSTM \\ MLP \\ SVM \end{tabular} & \begin{tabular}[c]{@{}c@{}} Positive \\ Negative \end{tabular} & 76.4\% of accuracy \\
    \hline
    
    \cite{piwowarski2019cognitive}& 2019 & \begin{tabular}[c]{@{}c@{}} EEG \\ ECG \\ EDA  \end{tabular} &  30 & Audiovisual & Not specified & Not specified & \begin{tabular}[c]{@{}c@{}} AWI \\ MI \end{tabular} & \begin{tabular}[c]{@{}c@{}} Study the effectiveness\\ of social advertising in\\ the context of promoting a \\healthy lifestyle \end{tabular} & \begin{tabular}[c]{@{}c@{}} Methods \\ suitable \\ for study\end{tabular} \\
    \hline
    
    \cite{hsu2020neuromarketing}& 2020 & EEG &  20 & Audiovisual & Not specified & \begin{tabular}[c]{@{}c@{}} Bayesian Factor \\ Chi-square analysis \end{tabular} & Not specified & \begin{tabular}[c]{@{}c@{}} Change \\ of decision \end{tabular} & \begin{tabular}[c]{@{}c@{}} Subliminal messages did \\ have significant influence \end{tabular} \\
    \hline

    \cite{mateusz2020cognitive}& 2020 & \begin{tabular}[c]{@{}c@{}} EEG \\ ECG \\ EDA  \end{tabular} &  20 & Audiovisual & \begin{tabular}[c]{@{}c@{}} Bandpass filter \\  ICA\end{tabular} & \begin{tabular}[c]{@{}c@{}} GFP \\ IAF \end{tabular} & \begin{tabular}[c]{@{}c@{}} MI \\ AW \\ EI \end{tabular} & \begin{tabular}[c]{@{}c@{}} Redesign advertisements to be\\ shorter and have a\\ better impact \end{tabular} & \begin{tabular}[c]{@{}c@{}} Relevant fragments\\ can be identified\end{tabular} \\
    \hline

    \cite{khushaba2012choice}& 2012 & EEG &  12 & Image & \begin{tabular}[c]{@{}c@{}} Bandpass filter \\ ICA \end{tabular} & GLM & Mutual Information & \begin{tabular}[c]{@{}c@{}} Impact \\of image\end{tabular} & \begin{tabular}[c]{@{}c@{}} Colors, patterns, or their combinations \\played a significant role in the selection of a product\end{tabular} \\
    \hline
    
    \cite{aldayel2020deep}& 2013 & EEG &  15 & Image & \begin{tabular}[c]{@{}c@{}} Bandpass filter \\ FIR1 \end{tabular} & PSD & \begin{tabular}[c]{@{}c@{}} RF \\ kNN \\ SVM \\ DNN \end{tabular} & Reaction & \begin{tabular}[c]{@{}c@{}}  94\% of accuracy\\  92\% of accuracy\\  62\% of accuracy\\ 88\% of accuracy\end{tabular} \\
    \hline
    
    \cite{vecchiato2014neuroelectrical} & 2014 & \begin{tabular}[c]{@{}c@{}} EEG \\ ECG \\EDA \end{tabular} & 20 & Image & \begin{tabular}[c]{@{}c@{}}Bandpass filter\\ ICA \\ CAR\end{tabular} & PSD & Statistics & \begin{tabular}[c]{@{}c@{}}Vote \\ Dominance \\ Trustworthiness\end{tabular} & 50\% of accuracy\\
    \hline
    
    \cite{matukin2016toward} & 2016 & \begin{tabular}[c]{@{}c@{}}EEG \\ ET\end{tabular} & 40 & Image & Bandpass filter & FFT & Not specified & Improving the advertisement design & \begin{tabular}[c]{@{}c@{}}Guidelines to advertisers \end{tabular} \\
    \hline
    
    \cite{rakshit2016discriminating}& 2016 & EEG & 7 & Image & \begin{tabular}[c]{@{}c@{}} Bandpass filter \\ CAR \end{tabular}  & Welch Method &  \begin{tabular}[c]{@{}c@{}} SVM \\ BPTT \end{tabular} & \begin{tabular}[c]{@{}c@{}} Red \\ Yellow \\ Green \\ Blue \end{tabular} & \begin{tabular}[c]{@{}c@{}} 85.26\% of accuracy\\ 80.2\% of accuracy\\ 78.32\% of accuracy\\ 76.4\% of accuracy\end{tabular}\\
    \hline

    \cite{samsuri2016left}& 2016 & \begin{tabular}[c]{@{}c@{}} EEG \\ ET \end{tabular} &  15 & Image & \begin{tabular}[c]{@{}c@{}} Bandpass filter \\ Baseline correction \end{tabular} & \begin{tabular}[c]{@{}c@{}} P300 ERP \\ N100 \end{tabular} & Statistics & Attention level & \begin{tabular}[c]{@{}c@{}} ERP and the ET results were \\ inconsistent at stimulus\\ discrimination task  \end{tabular} \\
    \hline

    \cite{golnar2019application}& 2019 &  EEG & 16  & Image & \begin{tabular}[c]{@{}c@{}} Bandpass filter \\ ICA \end{tabular} & PSD & \begin{tabular}[c]{@{}c@{}} SVM \\ LDA \end{tabular} & \begin{tabular}[c]{@{}c@{}} Like vs Neutral \\ Dislike vs Neutral \\ Buy vs Neutral\end{tabular} & \begin{tabular}[c]{@{}c@{}}87.06\% of accuracy \\ 95.95\% of accuracy\\93.87\% of accuracy\end{tabular} \\
    \hline
    
    \bottomrule
  \end{tabular}}
\end{table*}


\subsection{Comparison between advertisements}

The last objective compares different advertisements belonging to the same market niche. There is an interest in comparing them as they share the same objective or advertised product. Following the structure of the previous sections, the present section analyzes articles that use audiovisual stimuli, those that only use visuals, and the rest of the works. These studies have been collected and summarized in \tablename~\ref{tab:comparison}, where the main characteristics of each one are shown.

The main objective proposed in this objective is predicting a favorite item in a set of articles. This has been done using visual stimuli, as they can describe a product quickly and concisely. Murugappan et al., \cite{murugappan2014wireless} intended to obtain the favorite product of each subject by using the EEG signal, along with features related to PSD, Spectral Energy (SE), and Spectral Centroid (SC). After this, they generated models based on ML as kNN; and DL, like PNN algorithms, obtaining up to 96.6\% accuracy in detecting the product with the highest user preference. Following a similar objective, Oon et al., \cite{oon2018analysis} employed the EEG signal and features from DFA, obtaining 67.4\% accuracy with the NN algorithm. The same approach was followed by Hakim et al., \cite{hakim2021machines}, where they used FBP and hemispheric symmetry based feature extraction and ML algorithms such as SVM, LR, kNN, and DT, obtaining a maximum of 68.51\% accuracy in a binary classification. Other authors studied whether it is possible to recognize the probability of purchasing a product through exposure to audiovisual stimuli. This recognition makes it possible to compare different campaigns employing the likelihood of buying the advertised product. Following the same objective, Dmochowski et al., \cite{dmochowski2014audience} developed a study based on EEG where, using features related to the ISC, they were able to obtain 66\% accuracy. Similarly, Gupta et al., \cite{gupta2017correlation} conducted a survey using STFT applied to the EEG signal and classification algorithms such as kNN and PNN. For the experimentation, advertisements for commercial soap brands such as Pears, Lux, Cinthol and Dove were shown. The results yielded by this methodology indicated that these algorithms could identify the changes caused by different stimuli.

Another objective in neuromarketing studies is predicting the interest or ``Liking'' that the advertisement generates in the users. The first study for this purpose was conducted by Soria et al., \cite{soria2015advertising}, which performed an analysis based on ANN from the raw data of the EEG signal, obtaining 68\% accuracy in detecting whether a given advertisement is liked or disliked by the user. In the same year, Boksem et al., \cite{boksem2015brain} extracted the EEG signal of 29 subjects and extracted information related to the frequency of the signal using the FFT algorithm. They then applied regression-based classification algorithms, concluding that these methods adequately solved this problem. Later, Guixeres et al., \cite{guixeres2017consumer} pursued the same objective but added ECG and ET biosignals to the historically used EEG. The information extracted from the EEG signal was based on power using PSD and GFP, while they calculated the AOI for the eye-tracking signal. With this information, an ANN-based model was created to recognize those products the user liked, obtaining 82.9\% accuracy. Finally, Alimardani et al., \cite{alimardani2021deep} conducted a similar experiment where they used EEG signals and added visual stimuli from images. Again, frequency-related features of the signal were obtained by PSD, and for classification, they used SVM, RF, LR algorithms, and a CNN network. This work gathered a maximum accuracy of 74.56\% between all the algorithms used.

Some advertisements generate certain emotions in users, such as happiness or nostalgia, which increases the possibility of purchasing. For this reason, predicting when a stimulus generates a specific emotion is very interesting for companies. In 2009, Ohme et al., \cite{ohme2009analysis}, focused on studying the emotions evoked by each advertisement. The authors relied on the EEG biosignal, and EMG and FR, two classic techniques for recognizing emotions from facial gestures. In the case of the EEG signal, FFT was applied to study the different frequency bands directly related to emotions. This data extracts statistics and corroborates that different emotional responses are obtained depending on the design of the advertisement. With a similar objective, Ural et al., \cite{ural2019wavelet} studied how the frequency bands behave but, in this case, added EEG, ECG, and EDA biosignals. Again, the results showed that different band responses are gathered depending on the design of the advertisement. 

Other authors propose the detection of states other than classical ones. Among these, Barnett et al., \cite{barnett2017ticket} studied how to predict scores between zero and ten from the correlation of data from Cross-Brain Correlation (CBC) using EEG, ECG, and EDA. Later, Hakim et al., \cite{hakim2018pathways} conducted a study to predict a binary decision for selecting the favorite product. They used only the EEG signal and features extracted from STFT, FBP, hemispheric asymmetry, and ISC. After that, they applied both ML and DL algorithms, obtaining a maximum accuracy of 68.51\%.

The next category of stimulus used is visual, which leaves aside the auditory dimension. Similar to the work done with audiovisuals, predicting the purchase preference between a set of advertisements is of great interest. Some of the first authors who followed this objective with visual stimuli were Baldo et al., \cite{baldo2015brain}, which conducted a research where they performed a series of experiments from the EEG signal, training a 1D linear classifier algorithm, achieving an 80\% accuracy. In this sense, Yadava et al., \cite{yadava2017analysis} studied this same problem, but with the difference of applying DWT to extract relevant information from the EEG signal and using the HMM algorithm to classify the purchase preference. With this algorithm selection, they increased the classification accuracy to 95.33\%. Furthermore, Amin et al., \cite{amin2020consumer} changed the classification algorithm to DT, reaching up to 93\% accuracy. Similar to the objective of determining the preference when purchasing a product, obtaining a probability for user purchase is also of great interest. For this, Khusaba et al., \cite{khushaba2012neuroscientific} employed the EEG signal along with features obtained from FFT and the MI algorithm, obtaining as a result that their proposed framework can recognize purchase intentions. Similarly, Telpaz et al., \cite{telpaz2015using} obtained different evoked potentials such as N200 from the EEG signal along with neural random utility model algorithms, reaching 60\% accuracy. Later, Garczarek \cite{garczarek2018eeg, garczarek2018explicit} and Huseynov et al., \cite{huseynov2019incorporating} increased the number of biosensors, from EEG only, up to four biosignals such as EEG, ECG, EDA, and ET. These signals were collected from 21 subjects, obtaining at most 65.8\% accuracy.

As with audiovisual stimuli, obtaining how much a user has liked an image is an essential point of interest. To obtain this liking, Yilmaz et al., \cite{yilmaz2014like} considered the EEG signal and the PSD technique to extract the information from the image. Finally, they applied a linear regression algorithm, obtaining a model that recognizes when a user likes an image. Years later, Alday et al., \cite{aldayel2021recognition} performed a similar experiment, varying the algorithms for extracting characteristics. In this case, they use DWT and Welch method, increasing the number of algorithms to four: DNN, SVM, RF, and kNN. The maximum accuracy obtained in this case is 87\%.

Finally, new stimuli are proposed in addition to those mentioned above. One of these stimuli was used by Horska et al., \cite{horska2016innovative}, which employed the taste of different products to compare them. As biosignals, it used EEG and FR, obtaining statistics to predict whether it is a positive or negative stimulus. From the biosignal of FR, the authors extracted the facial expression from the user, which is related to emotion. The results show that these biosignals can detect the subjects' reactions, possibly predicting them. Bettiga et al., \cite{bettiga2020consumers} employed authentic products to study the subjects' reactions. The biosignals used were EEG, ECG, and EDA, obtaining the attention and pleasure indices from these biosignals. Correlation algorithms were then applied for classification, whose results show that it is possible to predict these reactions.

\begin{table*}[!ht]
  \caption{Works that compare campaigns.}
  \label{tab:comparison}
  \resizebox{\textwidth}{!}{
  \begin{tabular}{@{}cccccccccc@{}}
    \toprule
    \textbf{Ref} & \textbf{Year} &  \textbf{Biosignal} & \textbf{Subjects} &\textbf{Stimuli} & \textbf{Preprocessing} & \begin{tabular}[c]{@{}c@{}}\textbf{Feature} \\ \textbf{Extraction}\end{tabular} & \begin{tabular}[c]{@{}c@{}}\textbf{Classification} \end{tabular} & \begin{tabular}[c]{@{}c@{}}\textbf{Predicted} \\ \textbf{States}\end{tabular} & \textbf{Results}\\
    \midrule
    \hline
    
    \cite{ohme2009analysis}& 2009 & \begin{tabular}[c]{@{}c@{}} EEG \\ EMG \\ FR\end{tabular} & 45 & Audiovisual & \begin{tabular}[c]{@{}c@{}} Bandpass filter \\ Notch filter \\ ICA\end{tabular} & FFT & Statistics & \begin{tabular}[c]{@{}c@{}} Response \\ Emotion \end{tabular}  & \begin{tabular}[c]{@{}c@{}} Different responses \\ between advertisements \end{tabular} \\
    \hline

    \cite{murugappan2014wireless}& 2014 & EEG & 12 & Audiovisual & \begin{tabular}[c]{@{}c@{}} Bandpass filter\\ Surface Laplacian filter \end{tabular} & \begin{tabular}[c]{@{}c@{}} PSD\\ SE \\SC \end{tabular} & \begin{tabular}[c]{@{}c@{}} kNN\\ PNN \end{tabular} & Preference  & 96.62\% of accuracy \\
    \hline
    
    \cite{dmochowski2014audience}& 2014 & \begin{tabular}[c]{@{}c@{}} EEG\end{tabular} & 12 & Audiovisual & \begin{tabular}[c]{@{}c@{}} Bandpass filter \end{tabular} & \begin{tabular}[c]{@{}c@{}} ISC\end{tabular} & \begin{tabular}[c]{@{}c@{}} RA \end{tabular} & \begin{tabular}[c]{@{}c@{}} Purchase \\ intention \end{tabular} & \begin{tabular}[c]{@{}c@{}} 66\% of accuracy \end{tabular}\\
    \hline
        
    \cite{soria2015advertising}& 2015 & EEG & 10 & Audiovisual & Not specified & Not specified & ANN & Liking & 82\% of accuracy\\
    \hline
    
    \cite{boksem2015brain}& 2015 & EEG & 29 & Audiovisual & \begin{tabular}[c]{@{}c@{}} Downsampling \\Highpass \end{tabular} & FFT & \begin{tabular}[c]{@{}c@{}} Mix-Models \\ Regression \end{tabular} & Liking & \begin{tabular}[c]{@{}c@{}} EEG allows measurement\\ of purchase preferences \end{tabular} \\
    \hline
    
    \cite{barnett2017ticket}& 2017 & \begin{tabular}[c]{@{}c@{}} EEG\\ ECG \\ EDA \end{tabular} & 58 & Audiovisual & Not specified & CBC & Correlation & \begin{tabular}[c]{@{}c@{}} Rating \\ 0-10 \end{tabular} & 68\% of accuracy\\
    \hline
    
    \cite{christoforou2017your}& 2017 & \begin{tabular}[c]{@{}c@{}} EEG \\ ET \end{tabular} & 27 & Audiovisual & \begin{tabular}[c]{@{}c@{}} Downsampling \\ Highpass filter \\Notch filter \end{tabular} & \begin{tabular}[c]{@{}c@{}} Spectro-Temporal Domain \\ Cognitive-Congruency\end{tabular} & R2 & \begin{tabular}[c]{@{}c@{}} Variance \end{tabular} & 73\% of accuracy\\
    \hline
    
    \cite{guixeres2017consumer}& 2017 & \begin{tabular}[c]{@{}c@{}} EEG\\ ECG \\ ET \end{tabular} & 35 & Audiovisual & \begin{tabular}[c]{@{}c@{}} Downsampling\\ Bandpass filter \\ ICA \\ Interpolation \end{tabular} & \begin{tabular}[c]{@{}c@{}} PSD\\ GFP \end{tabular} & ANN & \begin{tabular}[c]{@{}c@{}} Liking \end{tabular} & 82.9\% of accuracy\\
    \hline
    
    \cite{gupta2017correlation}& 2017 & \begin{tabular}[c]{@{}c@{}} EEG\end{tabular} & 18 & Audiovisual & \begin{tabular}[c]{@{}c@{}} Bandpass filter \\ Notch filter \\Surface Laplacian filter \end{tabular} & \begin{tabular}[c]{@{}c@{}} STFT\end{tabular} & \begin{tabular}[c]{@{}c@{}} kNN \\ PNN \end{tabular} & \begin{tabular}[c]{@{}c@{}} Purchase \\ intention \end{tabular} & \begin{tabular}[c]{@{}c@{}} Dove brand advertisements\\ generate a greater impact\\ on subjects \end{tabular}\\
    \hline
    
    \cite{oon2018analysis}& 2018 & EEG & 10 & Audiovisual & \begin{tabular}[c]{@{}c@{}} Notch filter\\ Bandpass filter \end{tabular} & DFA & \begin{tabular}[c]{@{}c@{}} NN\\ kNN \end{tabular} & \begin{tabular}[c]{@{}c@{}} Preference \end{tabular}  & \begin{tabular}[c]{@{}c@{}} 67.44\% of accuracy \end{tabular} \\
    \hline
    
    \cite{hakim2018pathways}& 2018 & EEG & 33 & Audiovisual & \begin{tabular}[c]{@{}c@{}} Highpass filter\\ Notch filter \\ ICA \end{tabular} & \begin{tabular}[c]{@{}c@{}} STFT\\ FBP \\Hemisperic Asymmetry \\ ISC\end{tabular} & \begin{tabular}[c]{@{}c@{}} SVM\\ RGT\\ kNN \\ KDL \end{tabular} & \begin{tabular}[c]{@{}c@{}} Binary \\ choice \end{tabular} & 68.50\% of accuracy\\
    \hline
    
    \cite{ural2019wavelet}& 2019 & \begin{tabular}[c]{@{}c@{}} EEG \\ ECG \\ EDA\end{tabular} & 30 & Audiovisual & CWT & \begin{tabular}[c]{@{}c@{}} Wavelet Coherence \\ Phase Difference\end{tabular} & Statistics  & \begin{tabular}[c]{@{}c@{}} Alpha \\ Beta \\ Gamma \\ Delta \\ Teta \end{tabular}  & \begin{tabular}[c]{@{}c@{}} Coherence \\ in bands \end{tabular} \\
    \hline
    
    \cite{giroldini2020eeg}& 2020 & EEG & 23 & Audiovisual & Not specified & FFT & Not specified & \begin{tabular}[c]{@{}c@{}} Variations of \\ each band \end{tabular}  & \begin{tabular}[c]{@{}c@{}} Increase in EEG activity \end{tabular} \\
    \hline
    
    \cite{hakim2021machines}& 2021 & EEG & 13 & Audiovisual & \begin{tabular}[c]{@{}c@{}} Notch filter\\ ICA \end{tabular} & \begin{tabular}[c]{@{}c@{}} FBP\\ Hemispheric symmetry \end{tabular} & \begin{tabular}[c]{@{}c@{}} SVM\\ Logistic Regression \\ kNN \\ DT \end{tabular} & Preference  & 68.51\% of accuracy\\
    \hline

    \cite{alimardani2021deep}& 2021 & EEG & \begin{tabular}[c]{@{}c@{}} 25 \\ 35 \end{tabular} & \begin{tabular}[c]{@{}c@{}} Image \\ Audiovisual \end{tabular} & Bandpass filter & PSD & \begin{tabular}[c]{@{}c@{}} SVM \\ RF \\ Logistic Regression \\ CNN \end{tabular}& \begin{tabular}[c]{@{}c@{}} Liking \end{tabular}  & 74.57\% of accuracy \\
    \hline

    \cite{khushaba2012neuroscientific}& 2012 & \begin{tabular}[c]{@{}c@{}} EEG\end{tabular} & 18 & Image & \begin{tabular}[c]{@{}c@{}} Bandpass filter \\ PCA \end{tabular} & \begin{tabular}[c]{@{}c@{}} FFT\end{tabular} & MI & \begin{tabular}[c]{@{}c@{}} Purchase \\ intention \end{tabular} & \begin{tabular}[c]{@{}c@{}} Significant change in\\ spectral activities for\\ subject preferences
 \end{tabular}\\
    \hline
    
    \cite{khushaba2013consumer}& 2013 & \begin{tabular}[c]{@{}c@{}} EEG \\ ET \end{tabular} & 18 & Image & ICA & DWT & \begin{tabular}[c]{@{}c@{}} Mutual \\ Information \end{tabular} & \begin{tabular}[c]{@{}c@{}} Preference \end{tabular} & \begin{tabular}[c]{@{}c@{}} Changes in EEG \\ for preference \end{tabular} \\
    \hline
    
    \cite{yilmaz2014like}& 2014 & EEG & 15 & Image & \begin{tabular}[c]{@{}c@{}} Bandpass filter \\ Remove eye blink manually\end{tabular} & PSD & \begin{tabular}[c]{@{}c@{}} Logistic Regression \end{tabular}& \begin{tabular}[c]{@{}c@{}} Liking \end{tabular}  & \begin{tabular}[c]{@{}c@{}} Male and female behavior\\ for this set of stimulant\\ images were similar \end{tabular} \\
    \hline

    \cite{telpaz2015using}& 2015 & \begin{tabular}[c]{@{}c@{}} EEG\end{tabular} & 15 & Image & \begin{tabular}[c]{@{}c@{}} Notch filter \\ ICA \end{tabular} & \begin{tabular}[c]{@{}c@{}} N200 \\ Theta band power\end{tabular} & \begin{tabular}[c]{@{}c@{}} Neural random \\ utility model \end{tabular} & \begin{tabular}[c]{@{}c@{}} Purchase \\ intention \end{tabular} & \begin{tabular}[c]{@{}c@{}} 60\% of accuracy \end{tabular}\\
    \hline
    
     \cite{baldo2015brain}& 2015 & EEG & 40 & Image & \begin{tabular}[c]{@{}c@{}} Preference Index \end{tabular} & Not specified & \begin{tabular}[c]{@{}c@{}} 1D Linear \\ Classifier \end{tabular} & Preference & 80\% of accuracy\\
    \hline
    
    \cite{yadava2017analysis}& 2017 & EEG & 25 & Image & Savitzky-Golay & DWT & HMM & Preference  & 95.33\% of accuracy\\
    \hline
    
    \cite{garczarek2018eeg}& 2018 & \begin{tabular}[c]{@{}c@{}} EEG \\ECG \\EDA\\ ET \end{tabular} & 21 & Image & \begin{tabular}[c]{@{}c@{}} Notch filter \\ Bandpass filter \end{tabular} & Not specified & WALD & \begin{tabular}[c]{@{}c@{}} Purcharse \\ intention \end{tabular} & 65.8\% of accuracy\\
    \hline
    
    \cite{garczarek2018explicit}& 2018 & \begin{tabular}[c]{@{}c@{}} EEG \\ ET \end{tabular} & 16 & Image & \begin{tabular}[c]{@{}c@{}} Downsampling \\ Bandpass filter \end{tabular} & \begin{tabular}[c]{@{}c@{}} DWT\end{tabular} & Not specified & \begin{tabular}[c]{@{}c@{}} Purchase \\ intention \end{tabular} & \begin{tabular}[c]{@{}c@{}} Methods suitable \\ for study \end{tabular}\\
    \hline
    
    \cite{huseynov2019incorporating}& 2019 & \begin{tabular}[c]{@{}c@{}} EEG \\FR\\ ET \end{tabular} & 119 & Image & \begin{tabular}[c]{@{}c@{}} Bandpass filter \\ Downsampling \end{tabular} & FFT & LASSO & \begin{tabular}[c]{@{}c@{}} Purcharse \\ intention \end{tabular} & 60\% of accuracy\\
    \hline

    \cite{amin2020consumer}& 2020 & EEG & 25 & Image & CAR & DWT & DT & Preference  & 93\% \\
    \hline
        
    \cite{shaari2019electroencephalography}& 2019 & EEG & 5 & Image & Not specified & FFT & Not specified & \begin{tabular}[c]{@{}c@{}} Decision\\ making \end{tabular}  & \begin{tabular}[c]{@{}c@{}} Right hemisphere \\ is the most activated \end{tabular} \\
    \hline
    
    \cite{aldayel2021recognition}& 2021 & EEG & 25 & Image & \begin{tabular}[c]{@{}c@{}} Downsampling \\ Bandpass filter \\ICA \\ Savitzky-Golay\end{tabular} & \begin{tabular}[c]{@{}c@{}} DWT \\ Welch method\end{tabular} & \begin{tabular}[c]{@{}c@{}} DNN \\ SVM \\ kNN \\ RF \end{tabular}& \begin{tabular}[c]{@{}c@{}} Liking \end{tabular}  & 87\% of accuracy\\
    \hline
    
    \cite{horska2016innovative}& 2016 & \begin{tabular}[c]{@{}c@{}} EEG \\FR \end{tabular} & 22 & Flavor & \begin{tabular}[c]{@{}c@{}} Not specified \end{tabular} & Facial expression & Statistics & \begin{tabular}[c]{@{}c@{}} Positive \\ Negative \end{tabular} & \begin{tabular}[c]{@{}c@{}} Neuromarketing can be a useful tool\\ that will help manufacturers\\ and sellers to offer products that\\ truly, and not only in appearance,\\ satisfy customers \end{tabular}\\
    \hline
    
    \cite{bettiga2020consumers}& 2020 & \begin{tabular}[c]{@{}c@{}} EEG \\ ECG \\EDA \end{tabular} & 21 & Real Product & \begin{tabular}[c]{@{}c@{}} Not specified \end{tabular} & \begin{tabular}[c]{@{}c@{}} Attention index \\ Pleasant index \end{tabular} & Correlation & \begin{tabular}[c]{@{}c@{}} Subjects are exposed to\\ both utilitarian and\\ hedonic products \end{tabular} & \begin{tabular}[c]{@{}c@{}} Functional and hedonic products\\ generate emotional responses\\ in consumers \end{tabular}\\
    \hline
    
    \bottomrule
  \end{tabular}}
\end{table*}

\section{What datasets are publicly available in neuromarketing, and what type of data include?}
\label{datasets}

This section reviews public datasets currently available for neuromarketing objectives. However, these datasets have yet to be designed with the specific purpose of being used for neuromarketing. Consequently, the datasets most commonly used in the literature are those aimed at detecting different states of the person. Many of these datasets are previously labeled with aspects of subjects' states, such as cognitive load or emotional states. For all these reasons, multiple works focused on neuromarketing have opted to use these datasets to avoid data dependency.

\tablename~\ref{tab:datasets} shows the collected datasets used in neuromarketing works, as well as others that can be used in neuromarketing domains. For each of the datasets, this section takes into consideration the following characteristics: (i) the year of publication, (ii) the types of biosignals they incorporate, (iii) the number of subjects studied, (iv) the size of the dataset, (v) the stimuli used and (vi) the labels employed. The datasets collected in this work are public and can be requested from the respective authors.

The first public dataset about mental states was created in 2009, named ``Imagined Emotion'' \cite{onton2009high}. This data\-/set contains EEG and ECG signals, where 31 subjects were asked to remember a specific past situation to evoke an emotion. At the end of the experiment, each subject completed a questionnaire to assess the intensity of the emotion sought to be evoked. The data is labeled with the valence value, which can be translated to a specific mental state. In total, it contains 34GB of data from EEG and ECG signals. 

Three years later, the DEAP dataset \cite{koelstra2011deap} was created and became one of the most used in emotional recognition. This dataset comprises EEG, EMG, EDA, and ET biosignals obtained from 32 subjects, having a total size of 2.9GB. The authors used music videos to stimulate the subjects and label the data. After presenting these stimuli, the subjects had to perform the self-assessment questionnaire known as Self-Assessment Manikin (SAM). SAM is a questionnaire based on valence and arousal, which can directly translate into different emotional states. Moreover, the dataset also provides liking and dominance values for data labeling, which can be very useful in studies related to neuromarketing. Similarly, the MAHNOB HCI dataset \cite{soleymani2011multimodal} uses a very similar technique maintaining all previous biosignals and adding, in this case, FR as a biosignal. In addition, it also modifies the stimuli presented, using videos and images instead of music videos. The number of subjects and dataset size are also very similar, with 30 subjects participating in the experiment and 2.5GB in total size.

Later, in 2015, the DECAF\cite{abadi2015decaf} and SEED\cite{zheng2015investigating} datasets were created. The DECAF dataset is similar to those presented above, although varying the stimulus used, in this case, pieces of movies. DECAF is one of the most extensive datasets, weighing about 300GB, and includes 30 participants. Meanwhile, the SEED dataset modifies the labeling of the data concerning the DECAF dataset, assigning a specific state to each movie clip, thus eliminating each person's factor. The SEED dataset has certain drawbacks since the selected clip may not cause the states for which it was designed. One year later, the ASCERTAIN dataset \cite{subramanian2016ascertain} was constructed, where the number of subjects studied was increased to 58, maintaining the EEG, ECG, EDA, and FR biosensors. This dataset uses visual stimuli, and the data labeling was again performed using the SAM system, reaching a size of 56MB.

Yadava et al., \cite{yadava2017analysis} created in 2017 the first dataset designed specifically for neuromarketing work. The stimuli used in this dataset are a series of images of commercial products, labeled as ``like'' or ``dislike''. The objective of the dataset is to decide whether the subject liked them. These images were shown to 25 subjects, obtaining the EEG signal of each one of them. In addition, the second version of the SEED dataset appeared the same year, known as SEED-IV \cite{8283814}. It follows the same methodology as its previous version, although the labels assigned to the data vary. The labels included in this dataset are \textit{Fear}, \textit{Sadness}, \textit{Happiness}, and \textit{Neutral}.

Then, in 2018, Google created a massive dataset with people's faces to identify a particular emotional state from facial expressions \cite{47657}. The most remarkable feature of this dataset is that it is composed of more than 156,000 faces labeled as \textit{Happiness}, \textit{Anger}, \textit{Surprise}, and \textit{Contempt}. Despite a large number of images, the size of the dataset is reduced to 200MB. The same year, the DREAM\-/ER \cite{katsigiannis2017dreamer}, and CASE \cite{sharma2019dataset} datasets appeared. Both datasets are similar in terms of the type of stimulus selected and the stimuli they use in their work. The difference between them is the type of biosignals used. The DREAMER dataset uses the EEG and ECG signals, while the CASE dataset suppresses the EEG signal and adds the ECG, ET, EMG, EDA, Respiration, and Temperature signals.

In 2020, the DEAR-MULSEMEDIA dataset focused on stimulating all human senses, thus employing odors, tastes, texture, sounds, and videos \cite {raheel2021dear}. At the same time, they were capturing EEG, EDA, and ET biosignals, labeling these data using the SAM questionnaire. In the same year, the dataset called K-EmoCon \cite{park2020k} appeared, whose peculiarity is that the stimulus used is based on personal contact, simulating a debate between two people. The data were labeled from multiple sources, including the interviewee, the interviewer, and a series of observers who decided based on the whole context. 32 subjects were selected for this experiment, capturing EEG, ECG, EDA, Temperature, and Acceleration signals. The present work considers this dataset because one of the main challenges in neuromarketing is the use of new stimuli leaving aside the classic ones, and this work takes it into account. One of these stimuli is how other people influence the purchase of a product, so the K-EmoCon dataset helps study this task. Finally, in 2020, Daly et al., \cite{daly2020eeg} created a dataset to study how auditory stimuli affected a person's state. For this, the EEG signal of 31 participants was captured, labeling the data utilizing the SAM questionnaire and stimulating these subjects with different sounds.

Finally, the most recent dataset, MAUS, was released in 2021 \cite{beh2021maus}. It measures the cognitive load generated by a stimulus to a subject using the ECG, ET, and EDA signals. The stimuli used in this study is the N-Back task, which involves counting the number of digits that appear on the screen, following a series of patterns. For the creation of this dataset, 22 subjects participated, obtaining 77MB of data. The design of this dataset could help study how the arrangement of the objects and the patterns they follow affect their impact on the subjects.

\begin{table}[!ht]
  \caption{Public datasets available.}
  \label{tab:datasets}
  \centering
 \resizebox{0.48\textwidth}{!}{
  \begin{tabular}{@{}cccccccc@{}}
    \toprule
    \textbf{Ref} &\textbf{Name} & \textbf{Year} &  \textbf{Biosignals} & \textbf{Subjects}  & \textbf{Size} & \begin{tabular}[c]{@{}c@{}}\textbf{Stimuli} \\ \textbf{used}\end{tabular} & \begin{tabular}[c]{@{}c@{}}\textbf{Labels} \\ \textbf{used}\end{tabular}\\
    \midrule
    \hline
    
    \cite{onton2009high}& \begin{tabular}[c]{@{}c@{}} Imagined \\ emotion\end{tabular} & 2009 & \begin{tabular}[c]{@{}c@{}} EEG, ECG \end{tabular} & 31 & 34GB & \begin{tabular}[c]{@{}c@{}} Recollection of \\ past emotional \\ situations\end{tabular} & Valence \\
    \hline
    
    \cite{koelstra2011deap}& \begin{tabular}[c]{@{}c@{}} DEAP\end{tabular} & 2012 & \begin{tabular}[c]{@{}c@{}} EEG, EMG, \\ EDA, ET, \\ BVP, \\ Temperature\end{tabular} & 32 & 2.9GB & \begin{tabular}[c]{@{}c@{}} Music \\ videos\end{tabular} & SAM \\
    \hline
    
    \cite{soleymani2011multimodal}& \begin{tabular}[c]{@{}c@{}} MAHNOB HCI\end{tabular} & 2012 & \begin{tabular}[c]{@{}c@{}} EEG, ECG, \\ EDA, ET, \\ FR, \\ Temperature\end{tabular} & 30 & 2.5GB & \begin{tabular}[c]{@{}c@{}} Video \\ Images\end{tabular} & SAM \\
    \hline
    
    \cite{abadi2015decaf}& \begin{tabular}[c]{@{}c@{}} DECAF\end{tabular} & 2015 & \begin{tabular}[c]{@{}c@{}} EMG, MEG, \\ ECG, ET, \\ FR\end{tabular} & 30 & 300GB & \begin{tabular}[c]{@{}c@{}} Movie \\ clips\end{tabular} & SAM \\
    \hline
    
    \cite{zheng2015investigating}& \begin{tabular}[c]{@{}c@{}} SEED\end{tabular} & 2015 & \begin{tabular}[c]{@{}c@{}} EEG, \\ ET\end{tabular} & 15 & 6.95GB
    & \begin{tabular}[c]{@{}c@{}} Movie \\ clips\end{tabular} & \begin{tabular}[c]{@{}c@{}} Positive \\ Neutral \\ Negative\end{tabular} \\
    \hline
    
    \cite{subramanian2016ascertain}& \begin{tabular}[c]{@{}c@{}} ASCERTAIN\end{tabular} & 2016 & \begin{tabular}[c]{@{}c@{}} EEG, ECG, \\ EDA, \\ FR\end{tabular} & 58 & 56MB & \begin{tabular}[c]{@{}c@{}} Video\end{tabular} & \begin{tabular}[c]{@{}c@{}} SAM\end{tabular} \\
    \hline
    
    \cite{yadava2017analysis}& \begin{tabular}[c]{@{}c@{}} Analysis of EEG \\  signal in neuromarketing\end{tabular} & 2017 & \begin{tabular}[c]{@{}c@{}} EEG \end{tabular} & 25 & 86MB & \begin{tabular}[c]{@{}c@{}} Images\end{tabular} & \begin{tabular}[c]{@{}c@{}} Like \\ Dislike\end{tabular} \\
    \hline
    
    \cite{8283814}& \begin{tabular}[c]{@{}c@{}} SEED IV\end{tabular} & 2017 & \begin{tabular}[c]{@{}c@{}} EEG, ET \end{tabular} & 15 & 10GB & \begin{tabular}[c]{@{}c@{}} Movie \\ clips \end{tabular} & \begin{tabular}[c]{@{}c@{}} Happy \\ Sad \\Fear \\ Neutral\end{tabular} \\
    \hline
    
    \cite{47657}& \begin{tabular}[c]{@{}c@{}} Google facial \\  expression comparison \\ dataset\end{tabular} & 2018 & \begin{tabular}[c]{@{}c@{}} FR \end{tabular} & 156000 & 200MB & \begin{tabular}[c]{@{}c@{}} N.A \end{tabular} & \begin{tabular}[c]{@{}c@{}} Happiness \\ Anger \\Surprise \\ Content\end{tabular} \\
    \hline
    
    \cite{katsigiannis2017dreamer}& \begin{tabular}[c]{@{}c@{}} DREAMER\end{tabular} & 2018 & \begin{tabular}[c]{@{}c@{}} EEG, \\ ECG \end{tabular} & 23 & 500MB & \begin{tabular}[c]{@{}c@{}} Audiovisual \end{tabular} & \begin{tabular}[c]{@{}c@{}} SAM \end{tabular} \\
    \hline
    
    \cite{sharma2019dataset}& \begin{tabular}[c]{@{}c@{}} CASE\end{tabular} & 2018 & \begin{tabular}[c]{@{}c@{}} ECG, ET, \\  EMG, EDA, \\ Respiration, \\ Temperature \end{tabular} & 30 & 5GB & \begin{tabular}[c]{@{}c@{}} Audiovisual \end{tabular} & \begin{tabular}[c]{@{}c@{}} SAM \end{tabular} \\
    \hline
    
    \cite{raheel2021dear}& \begin{tabular}[c]{@{}c@{}} DEAR-MULSEMEDIA\end{tabular} & 2020 & \begin{tabular}[c]{@{}c@{}} EEG, EDA, \\ ET \end{tabular} & 18 & 1GB & \begin{tabular}[c]{@{}c@{}} Multiple \\ senses \end{tabular} & \begin{tabular}[c]{@{}c@{}} SAM \end{tabular} \\
    \hline

    \cite{park2020k}& \begin{tabular}[c]{@{}c@{}} K-EmoCon\end{tabular} & 2020 & \begin{tabular}[c]{@{}c@{}} EEG, ECG, \\EDA, Temperature \\ Acceleration\end{tabular} & 32 & 100MB & \begin{tabular}[c]{@{}c@{}} People-to-people \\ discussions \end{tabular} & \begin{tabular}[c]{@{}c@{}} Labeled from \\ three sources \end{tabular} \\
    \hline
    
    \cite{daly2020eeg}& \begin{tabular}[c]{@{}c@{}} An EEG dataset \\ recorded during affective \\  music listening\end{tabular} & 2020 & \begin{tabular}[c]{@{}c@{}} EEG \end{tabular} & 31 & 100MB & \begin{tabular}[c]{@{}c@{}} Auditory  \end{tabular} & \begin{tabular}[c]{@{}c@{}} SAM \end{tabular} \\
    \hline
    
    \cite{beh2021maus}& \begin{tabular}[c]{@{}c@{}} MAUS\end{tabular} & 2021 & \begin{tabular}[c]{@{}c@{}} ECG, PPG, \\ EDA\end{tabular} & 22 & 77MB & \begin{tabular}[c]{@{}c@{}} N-Back task  \end{tabular} & \begin{tabular}[c]{@{}c@{}} Cognitive \\ load \end{tabular} \\
    \hline
    
    \bottomrule
  \end{tabular}}
\end{table}

\section{How has the role of data fusion in neuromarketing evolved over time, and what are the open challenges?}
\label{ctl}
Based on the different aspects of neuromarketing studied together with the biosignals analyzed in questions \textit{Q1-Q3}, this section answers question \textit{Q4}. To solve this question, lessons learned, trends, and open challenges drawn from current work on neuromarketing are discussed.

\subsection{Lessons learned}
\label{lessons}

After reviewing the state of the art, the following lessons learned are summarized: 

\begin{itemize}

\item \textbf{Improving results through data fusion}. Effective data fusion techniques can enhance the accuracy and reliability of neuromarketing studies by combining multiple data modalities, such as EEG signals, eye tracking data, and physiological measurements. This integration allows for a more comprehensive understanding of consumer responses and behaviors.

\item \textbf{The comparison between advertisements is the most commonly used neuromarketing objective}. As shown in \figurename~\ref{fig:lifecycle}, almost 45\% of the works focus on comparing objectives. Its relevance is mainly because different products compete against each other to get the most prominent market quota, so obtaining feedback about the competition is very interesting for companies and, therefore, for research. However, the distribution of works in the rest of the objectives is quite balanced, being 30.4\% for the study of an advertisement and 24.8\% for the improvement of an advertisement.

\item \textbf{There is a large number of works following the visual-statistical method}. Many works in the literature used a classification method, collecting statistics from the data, then presenting the results visually. This technique is between 9-20\% among the different objectives, with the improvement objective being the one where it is used the most. This type of classification is very vulnerable to the subjectivity of the person studying the results, which is why ML/DL algorithms are more advisable in these situations.

\item \textbf{Audiovisual is the most common stimulus in neuromarketing}. These inputs are used in television advertisements, thus perpetuating this kind of stimulus. This category is the most used in the three objectives, exceeding 50\% of use in all of them (see \figurename~\ref{fig:disstimuli}). Nowadays, with the popularity of smartphones and tablets, advertisements are still audiovisual, although they add a degree of interaction with the user, which must be studied. After the audiovisual stimuli, some are solely visual through images. Following the same theory as audiovisual stimuli, images shown on television, billboards, and newspapers are the most common way to show a product to a potential consumer. Those works that use more innovative stimuli remain in the background, such as web pages or authentic products for their promotion.

\item \textbf{EEG is the most widespread biosignal in neuromarketing}. As seen in \figurename~\ref{fig:lifecycle}, the EEG biosignal is used in almost 50\% of the works studied. EEG is considered one of the most representative signals for identifying a subject's mental state. \figurename~\ref{fig:disbiosignals} shows the distribution of biosignals in each objective. In objectives aiming to provide a comparison or the improvement of advertisements, the use of EEG exceeds 50\%, while in the objective focused on the study of the impact of a stimulus, a homogeneous distribution between EEG, ET, and EDA signals is maintained, being EDA and ET the second most used signals. The EDA biosignal allows the identification of subjects' stress from the conductivity of the skin. In the case of the ET signal, it is mainly used in studies that incorporate visual stimuli. In contrast, ET provides visualization patterns or specific points where the user's attention has been focused. Employing these characteristics, it is possible to redesign a stimulus or verify if it fulfills the function for which it has been designed. 

\item \textbf{There are no unified data preprocessing techniques for neuromarketing objectives}. In almost all the studied works, the data processing techniques are different. \figurename~\ref{lifecycle} shows how the preprocessing and feature extraction techniques are very varied. Some patterns exist, such as using algorithms based on power and frequency to be applied to the EEG signal or the AOI for ET signals. Nevertheless, their combinations and the algorithms used for each pattern vary considerably. In addition, the results are data-dependent, so the definition of a unified set of algorithms for all use cases is an open challenge. 

\item \textbf{Public datasets are scarce and are not designed for neuromarketing objectives}. The public datasets that can be used in this type of work are very scarce (see \tablename~\ref{tab:datasets}). These datasets are mainly aimed at mental state detection, most of which have been created using stimuli to evoke emotional states. Most of the studied works do not provide the data publicly, unable to replicate the experiments performed.

\end{itemize}

\subsection{Current trends}
\label{trends}

\begin{figure*} [!ht]
     \begin{subfigure}[b]{.49\textwidth}
         \centering
         \includegraphics[width=\columnwidth]{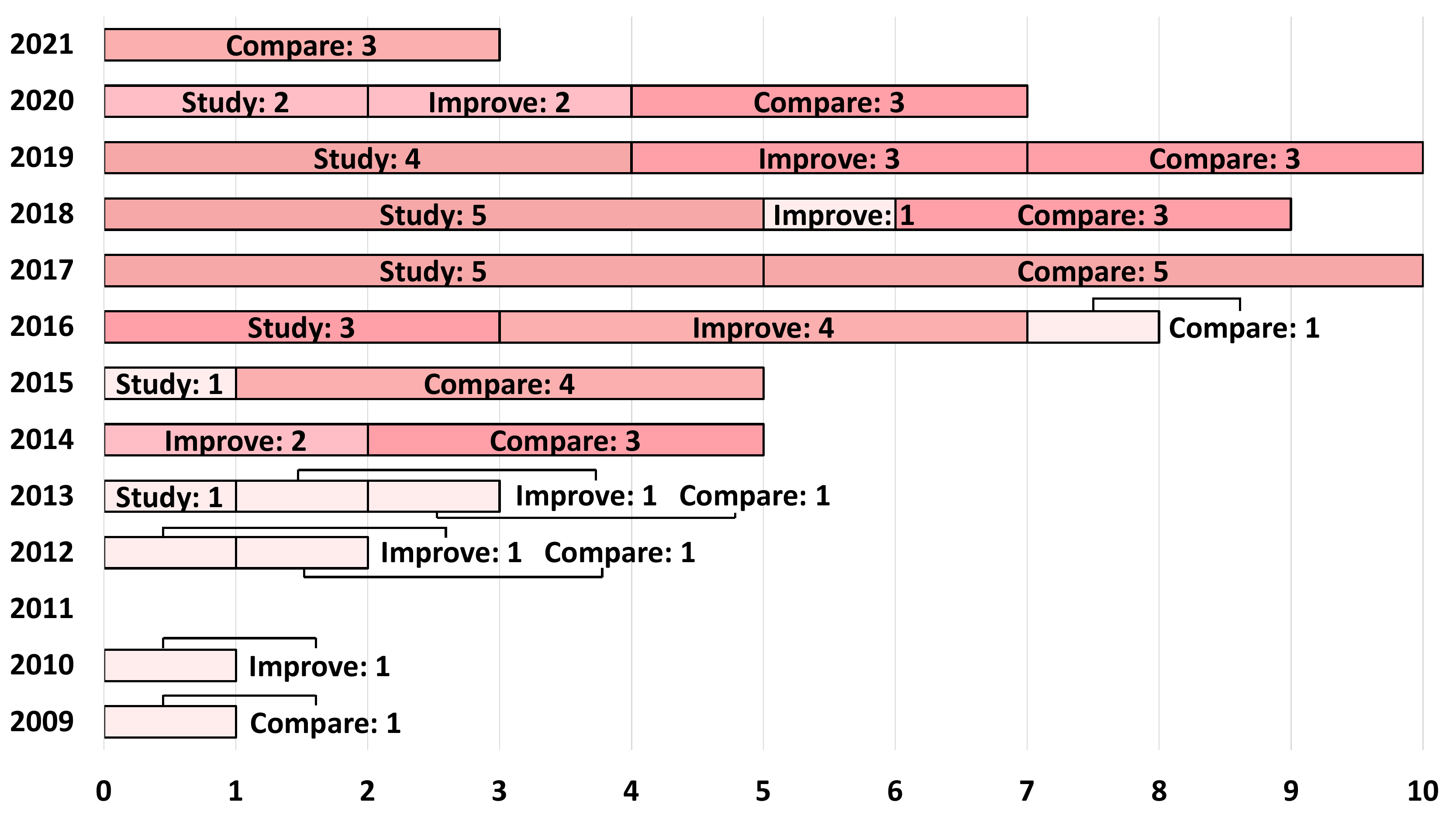}
         \caption{Number of works using each objective.}
         \label{fig:trendstrabajos}
     \end{subfigure}
      \begin{subfigure}[b]{.49\textwidth}
         \centering
         \includegraphics[width=\columnwidth]{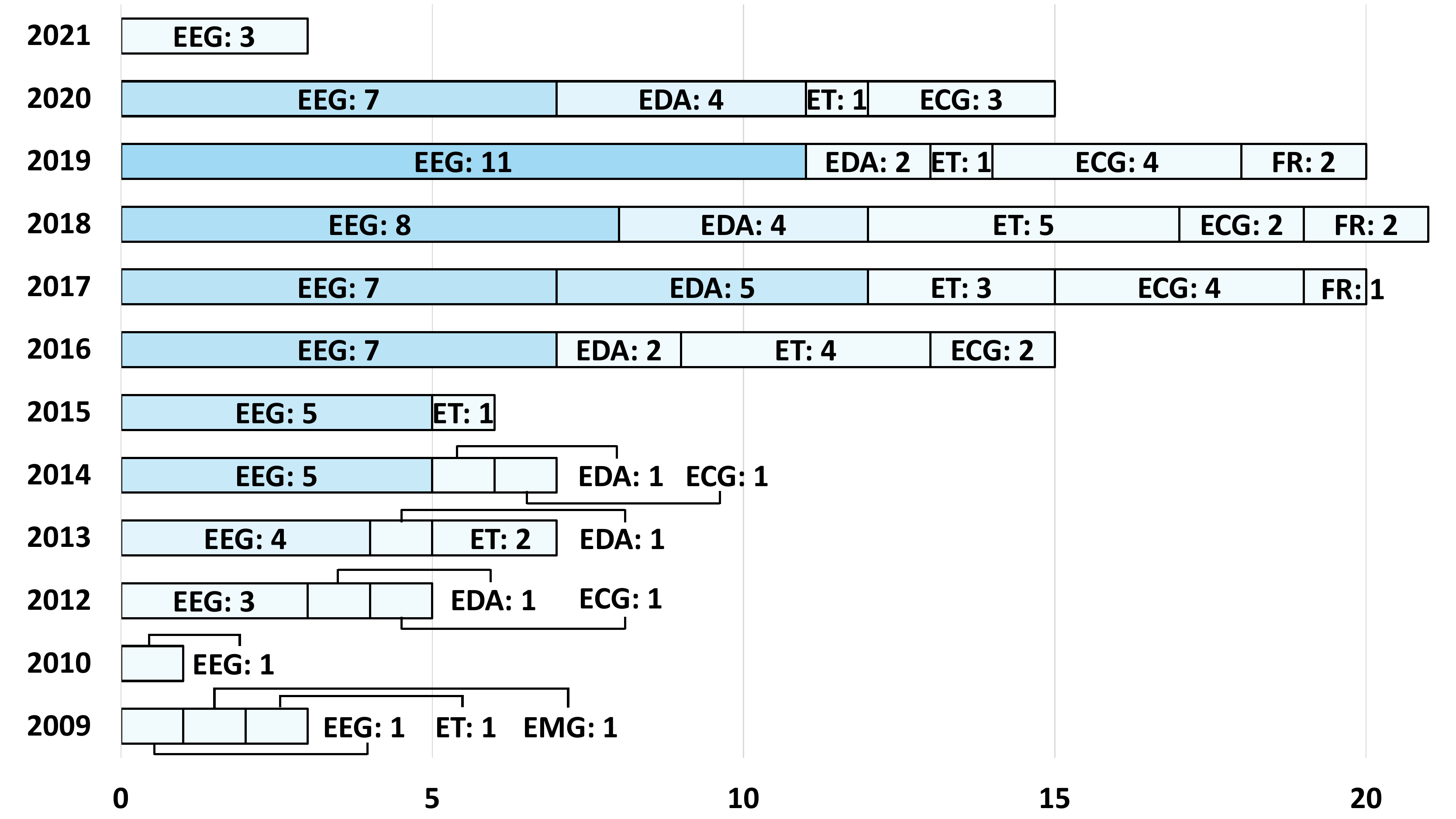}
         \caption{Number of works using each biosignal.}
         \label{fig:trendsSignals}
     \end{subfigure}
  
    \begin{subfigure}[b]{.49\textwidth}
         \centering
         \includegraphics[width=\columnwidth]{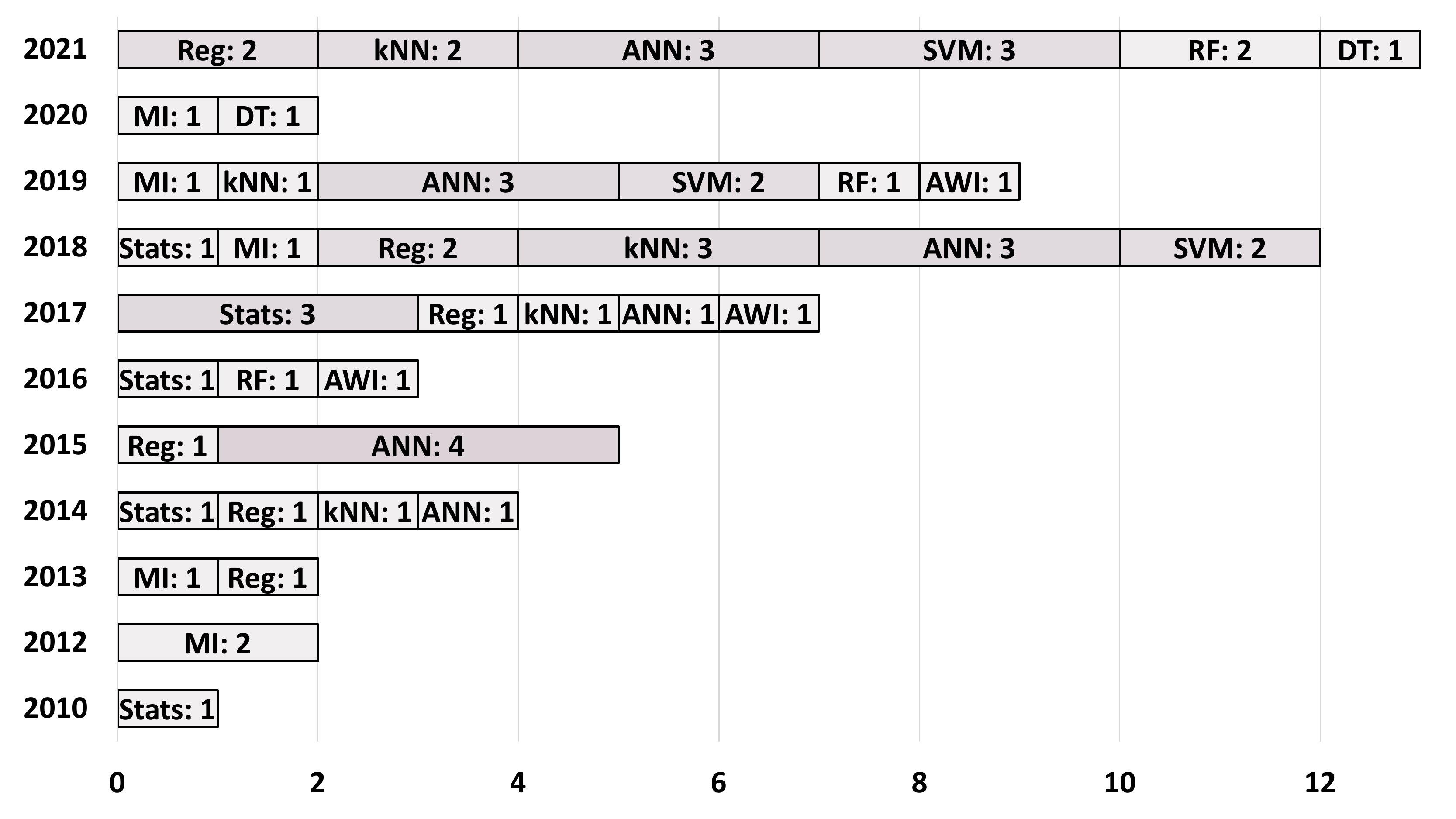}
         \caption{Number of works using each tool.}
         \label{fig:trendsML}
    \end{subfigure}
      \begin{subfigure}[b]{.49\textwidth}
         \centering
         \includegraphics[width=\columnwidth]{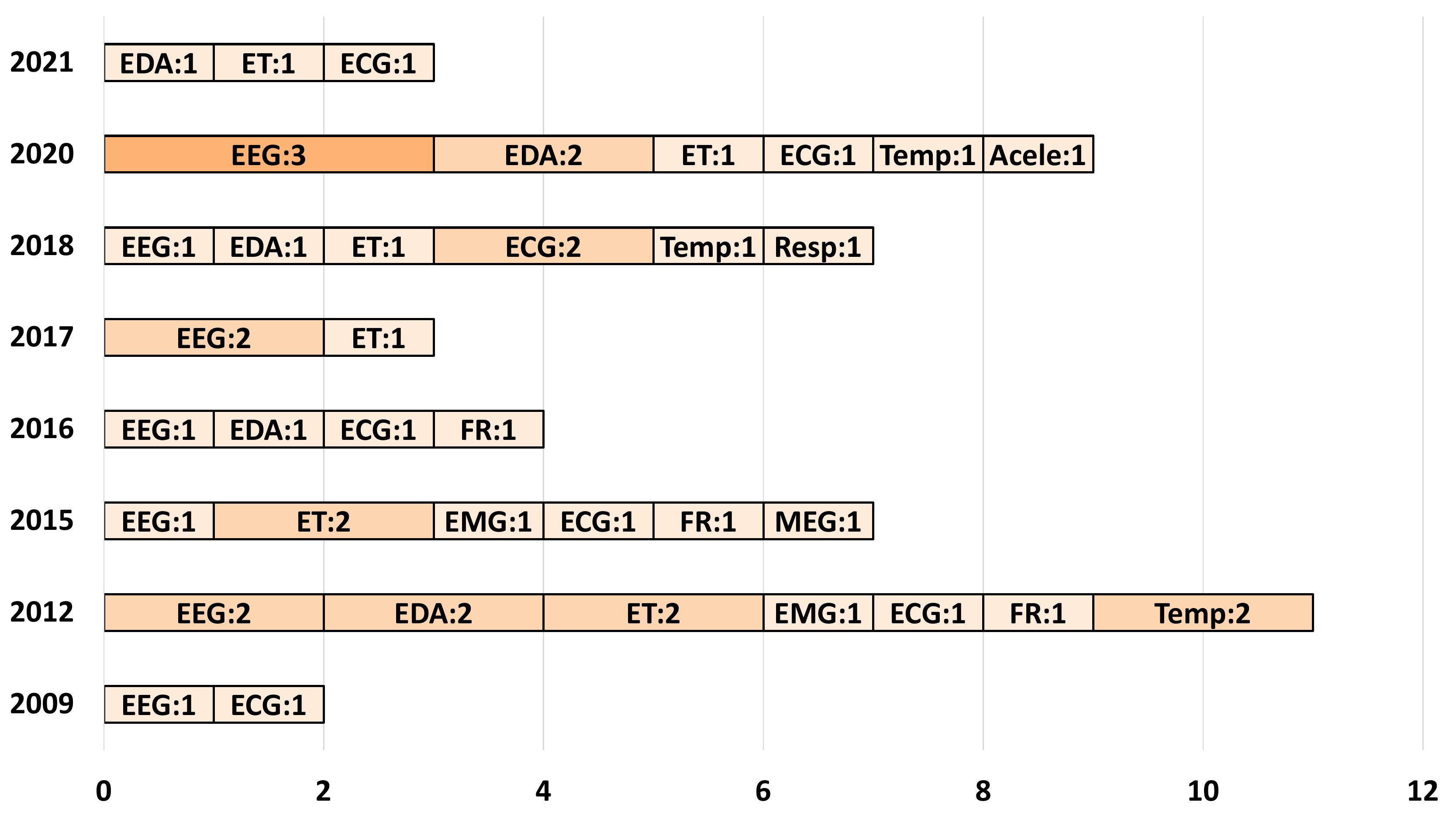}
         \caption{Number of datasets using each biosignal.}
         \label{fig:trendsdatasets}
    \end{subfigure}

  \caption{Trends in the works and datasets studied per year.}
  \label{fig:trends}
\end{figure*}

The main trends that can be drawn from the temporal analysis of the published papers and datasets are as follows:

\begin{itemize}
    \item \textbf{The study of advertisements has gained relevance in recent years}. \figurename~\ref{fig:trendstrabajos} shows how the advertisement comparison objective has been the most commonly proposed. However, in recent years, the study of advertisement has gained relevance, being the most employed in 2017, 2018, and 2019. This is because the possibilities for innovation within this objective are more significant than the rest.

    \item \textbf{The combination of EEG with other biosignals has grown in recent years}. The EEG signal has historically been the most commonly used. However, this signal difficult the interpretation of the data. For this reason, the literature has incorporated the use of other biosignals. Specifically, \figurename~\ref{fig:trendsSignals} shows how since 2016, the use of biosignals such as EDA or ECG has increased dramatically. Also, ET has been one of the most used for comparing and studying a particular advertisement, mainly because the primary stimulus used in these works is visual, thus having a close relationship.

    \item \textbf{ML and DL algorithms have gained relevance in recent years}. The emergence of ML and DL algorithms has created a new opportunity for automatic pattern identification in recent years. Historically, this was done by extracting specific features from relevant data and subjecting them to the human eye, but as shown in \figurename~\ref{fig:trendsML}, ML and DL algorithms have gained prominence in recent years, becoming the most widely used.

    \item \textbf{Datasets have not evolved much}. \figurename~\ref{fig:trendsdatasets} shows how the public datasets analyzed in this work have not undergone many modifications in biosignals, subjects, or stimuli used since their appearance. In these datasets, the EEG signal continues to predominate, followed only by the ET signal. 

\end{itemize}

\subsection{Challenges}
\label{challenges}

Based on the study of state of the art, this survey defines below the future challenges for further research that should be taken into account:

\begin{itemize}

    \item \textbf{Unified preprocessing techniques for multiple data sources}. One of the major challenges in data fusion for neuromarketing is the integration of heterogeneous data sources. Different data modalities may have varying sampling rates, noise levels, and feature representations, requiring careful preprocessing and alignment to ensure accurate fusion and interpretation of results.

    \item \textbf{Increase the use of stimuli that resemble those found in real use cases}. The use of stimuli based on sounds and images has been used in more than 90\% of the studies, as can be seen in \figurename~\ref{lifecycle}. Regarding neuromarketing, there is a wide variety of stimuli that can be used to modify a subject's attitude towards a product, such as dealing with other people when shopping or arranging objects in a store \cite{bastiaansen2018my}. However, these stimuli have not been exploited and studied in most works. This is the reason for the emergence of concepts such as Nanomarketing \cite{mileti2016nanomarketing}, which aims to study dimensions beyond visual an auditory inputs, such as the arrangement of objects in a store or interacting with the seller.

    \item \textbf{Increased use of cheaper and simpler biosensors allowing real-time response}. One of the most commonly used biosensors is the EEG signal (see \figurename~\ref{fig:disbiosignals}). However, although technology is evolving towards more portable and comfortable devices, the precision these devices can achieve is still reduced. Therefore, biosignals such as EDA and ECG should be considered in situations intended to identify more forthright states, such as stress, rather than an emotional state. These devices report a value easily interpreted by a machine, identifying trivial subject states.

    \item \textbf{Need for data labeling independent of users' subjectivity}. The SAM method is one of the most widely used approaches for labeling data. This method is based on obtaining specific parameters such as valence and arousal from the subject. It is a subjective evaluation that can lead to labeling and classification errors. Due to these errors, it is necessary to develop other methods for labeling data that are more accurate and leave aside the user's personal opinion.

    \item \textbf{The use and proposal of new and more descriptive features for classification}. As shown in \figurename~\ref{lifecycle}, only sensor biosignals are used as inputs to the classification algorithm. However, other information, such as stimulus-related features, can be used for classification, helping to obtain better classification results.

    \item \textbf{Use of models that ensure the privacy of the user's data.} When publishing public datasets, it is necessary to consider certain aspects regarding the privacy of these data. Because the information obtained from these devices has medical grade, specific private data such as purchase priorities may be extracted. For this reason, it is advisable to anonymize the data and link them only to the information necessary for the study, such as age, gender, or mental health conditions.
    
    \item \textbf{The optimization of ML and DL models to improve training speed.} The AI models created for biosignal classification have a high training cost because the variability of biosignals is large, so a large amount of data is needed. For this reason, it is necessary to investigate new training systems to improve performance in real use cases. Some techniques, such as transfer learning, have been tested in biosignals like EEG, ECG, or FR, which could help to solve this problem. In another direction, the possibility of training a model with a small group of users, which is generally applicable to all users, should be investigated \cite{wan2021review, jang2021effectiveness, podder2022time}. This enables direct application in a real use case without requiring specific training per user.
       
     \item \textbf{The creation of more public datasets designed for neuromarketing is needed}. As described in Section \ref{datasets}, the existence of public datasets created based on neuromarketing stimuli is reduced to one. Because of this, the generation of stimuli is not comparable to real-life neuromarketing scenarios, which may cause the methodology designed for that theoretical use case to be inapplicable in a real scenario. In addition, a more comprehensive range of public datasets can help use these datasets for new work, reducing the reliance on data when developing a methodology. In these common data, it is possible to study the proposed methodologies more fairly.

\end{itemize}

\section{Conclusions and future work}
\label{conclusions}
Neuromarketing studies have historical importance due to the impact of advertisements on companies sales and profits. However, the techniques used in these studies are very heterogeneous and it is crucial to know which of these techniques are the most appropriate depending on the neuromarketing objective. In order to fulfill this necessity, some surveys compile all this information and concisely indicate the main aspects of each neuromarketing phase. Although these works are useful, there are some shortcomings that still need to be solved. On the one hand, existing surveys need to consider the whole neuromarketing cycle, not leaving aside the technical part. Due to this lack of information, it is difficult to repeat the experimentation of a given study or to apply the methodology proposed by that study in another scenario. On the other hand, the techniques used in these studies are very heterogeneous, being necessary to know which techniques best fit the requirements depending on the neuromarketing objective. To improve these limitations, this paper reviews the phases of the neuromarketing cycle and analyzes the techniques used in the literature. In addition, to facilitate the understanding and selection of techniques, the works analyzed are divided according to the objective for which they have been designed. To achieve these objectives, the following research questions have been answered:

\textit{Q1. What are the most applied objectives, biosensors, and processing techniques in neuromarketing?} This work shows in a summarized form the most used techniques in each phase of the neuromarketing cycle. For the first phase, focused on the neuromarketing objectives, a predominance of the advertisement comparison objective is observed, while audiovisual stimuli predominate in the second stage. In later phases, the most used biosignal in neuromarketing is EEG, directly conditioning the data preprocessing techniques in subsequent steps. Since the processing is so heterogeneous, the summary of these procedures is complex. However, when conducting a neuromarketing study, one of the most relevant points is the techniques used to recognize a particular state. To get a better overview of these techniques, \figurename~\ref{fig:disml} shows the relationship of use for each of the objectives.

\textit{Q2. What combinations of data from biosignals, stimuli, and processing and classification techniques are suitable for each objective?} This work reviews which data fusions have been most commonly employed depending on the objective of the work. To better understand these relationships, \figurename~\ref{fig:relation} shows which stimuli have been the most used in each objective, as well as the biosignals and classification procedures. The results show the predominance of visual stimuli, EEG, and ML algorithms in most neuromarketing studies. In particular cases, the ET biosignal, and techniques such as heat mapping, become relevant for objectives aiming to study and improve advertisements.

\textit{Q3. What datasets are publicly available in neuromarketing, and what type of data include?} This paper reviews the datasets and the types of data they collect for neuromarketing studies, as well as those that can be applied due to their similarity. In conclusion, there is a need for datasets explicitly designed for neuromarketing studies. Consequently, most authors have used datasets whose labeling corresponds closely to the metrics used in a neuromarketing study. Moreover, the development trend among these datasets is minimal, implying that new datasets do not substantially expand upon or innovate beyond previously existing ones.

\textit{Q4. How has the role of data fusion in neuromarketing evolved over time, and what are the open challenges?} Historically, the data collected have been used to compare advertisements. However, in recent years, the study and improvement of a stimulus has gained relevance in the literature. The importance obtained by these objectives has caused new challenges related to neuromarketing. One of these challenges is the need for new datasets designed explicitly for neuromarketing and adjusted to the three objectives shown. In the same way, using stimuli that resemble the reality, such as real products or the arrangement of objects inside a store, is a potential need. ML/DL models usually recognize the states generated by these stimuli. Related to these models, the security they apply to users' data is a challenge that remains to be solved, as well as the performance of these when performing classification.

Aligned with current trends and challenges, future work could focus on designing and implementing new systems for predicting mental states in neuromarketing. In addition, there is an opportunity for proposing new stimuli and realistic scenarios for applying the designed techniques, as well as improvements in training and classification algorithms to make them applicable in real time. Finally, a dataset could be designed and made publicly available so that other researchers can use it to develop new research.

\section{Acknowledgments}

This work has been partially supported by (a) 21628/FPI/21 Fundación Séneca, cofunded by Bit \& Brain Technologies S.L., Región de Murcia (Spain), (b) the Swiss Federal Office for Defense Procurement (armasuisse) with the CyberTracer (CYD-C-2020003) project, and (c) the University of Zürich UZH.

\bibliography{references}

\end{document}